\documentclass[final,5p,twocolumn]{elsarticle}

\usepackage{amssymb}
\usepackage{lineno}

\usepackage{color}
\usepackage{booktabs}
\usepackage{makecell}
\usepackage{pifont}
\usepackage{rotating}
\usepackage{adjustbox}

\newcommand{\cmark}{\ding{51}}
\newcommand{\xmark}{\ding{55}}

\newcommand{\revision}[1]{{\color{black}{#1}}}

\journal{Computer \& Security}

\begin{document}
\begin{frontmatter}

\title{Threat Detection and Investigation with System-level Provenance Graphs: A Survey}

\author[label1]{Zhenyuan Li}
\author[label2]{Qi Alfred Chen}
\author[label1]{Runqing Yang}
\author[label3]{Yan Chen}
\author[label1]{Wei Ruan}

\address[label1]{Zhejiang University, China}
\address[label2]{University of California, Irvine, USA}
\address[label3]{Northwestern University, USA}

\begin{abstract}
With the development of information technology, the border of the cyberspace gets much broader and thus also exposes increasingly more vulnerabilities to attackers. Traditional mitigation-based defence strategies are challenging to cope with the current complicated situation. Security practitioners urgently need better tools to describe and modelling attacks for defense. 

The provenance graph seems like an ideal method for threat modelling with powerful semantic expression ability and attacks historic correlation ability. In this paper, we firstly introduce the basic concepts about system-level provenance graph and present a typical system architecture for provenance graph-based threat detection and investigation. A comprehensive provenance graph-based threat detection system can be divided into three modules: \textit{data collection module}, \textit{data management module}, and \textit{threat detection modules}. Each module contains several components and involves different research problems. We systematically taxonomize and compare the existing algorithms and designs involved in them. Based on these comparisons, we identify the strategy of technology selection for real-world deployment. We also provide insights and challenges about the existing work to guide future research in this area.
\end{abstract}


\end{frontmatter}


\section{Introduction}
\label{sec:introduction}

Threat detection in cyberspace is an arms race between adversaries and defenders. In this arms race, attackers can almost always bypass existing detection mechanisms by discovering new attack surfaces, while defenders are usually tired of plugging various vulnerabilities \cite{Symantec}. Therefore, it is necessary for security researchers and practitioners to start rethinking about traditional mitigation techniques and try to design more robust and general detection mechanisms against not only the various existing attacks but also the previously unseen ones. The Defense's Advanced Research Projects Agency (DARPA) has launched a four-year project called Transparent Computing since 2015 \cite{transparent-computing}, trying to find a high-fidelity and visible method to abstract the interaction between components in the opaque system. The researchers found that the provenance graph may be a promising tool, with a strong abstract expression ability and relatively high efficiency. 

\begin{figure}
    \centering
    \includegraphics[width=3.4in]{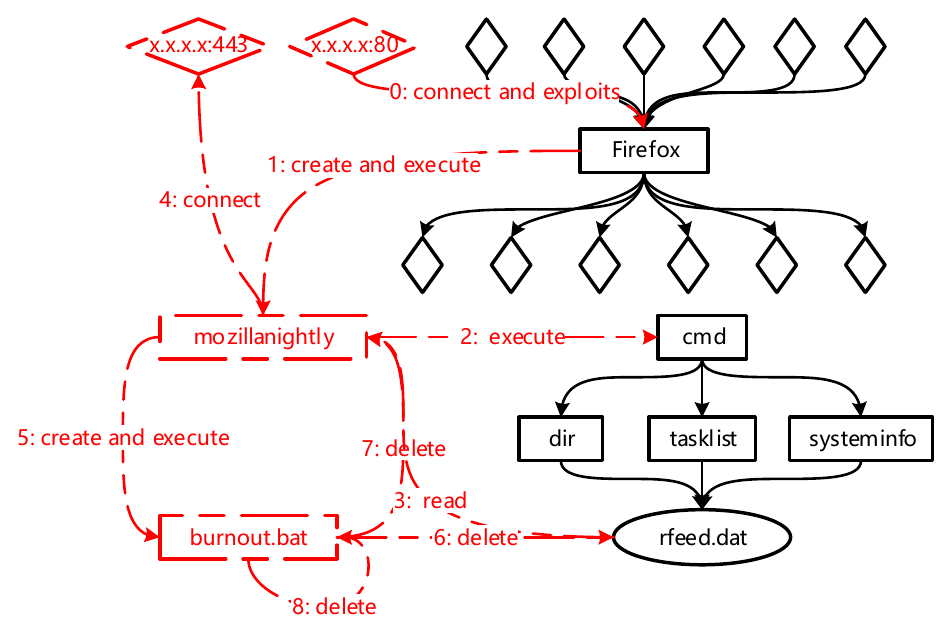}
    \caption{A provenance graph sample}
    \label{fig:samples}
\end{figure}

Now more and more research works \cite{hossain_sleuth:_2017, milajerdi_holmes_2019, xie_pagoda:_2018, milajerdi_poirot:_2019,hassan_nodoze:_2019,xie_p-gaussian_2019, ma_accurate_2015, barre2019mining} began to focus on detection and response algorithms based on provenance graphs and believe that provenance graph has the potential to become the next generation of more robust detection mechanisms. As shown in the Figure \ref{fig:samples}, the provenance graph represents the relationship between the control flow and data flow between the subject (such as processes, threads, etc.) and the object (such as files, registry, network sockets) in the system through a directed graph with timing. The provenance graph can link causal events in the system, regardless of the time between the two events. All in all, utilizing provenance graphs for threat detection and investigation has the following advantages:

\vspace{-\topsep}
\begin{list}{\labelitemi}{\leftmargin=1.5em}
 \setlength{\topmargin}{0pt}
 \setlength{\itemsep}{0em}
 \setlength{\parskip}{0pt}
 \setlength{\parsep}{0pt}
 
\item Provenance graphs altogether show system execution by representing them as interactions between system objects. Such dependency is innate for all the execution trace. Unstructured log like Auditd \cite{auditd} can also be transformed into provenance graph \cite{gehani_spade:_2012}.

\item Provenance graphs enable semantic-aware and robust detection. Compared to unstructured audit logs, provenance graphs with spatial and temporal information are more difficult to forge by attackers \cite{han2018provenance}. Moreover, provenance graphs provide richer semantic; thus security analysts can conduct more effective and thorough attack investigation.

\item Provenance graphs keep all the execution history. Advanced persistent threat (APT) attacks \cite{APT} are long-running and stealthy attacks. To investigate such attacks, analysts need to access and understand the whole attack history. Actually, system execution history is necessary for any intrusion to trace the entry point and understand the impact.
\end{list}

\revision{

To take advantage of the provenance graph, security researchers need to design and implement provenance graph-based detection systems. A typical system can be divided into three sub-modules: ``data collection module'' (\S\ref{sec:data_collection}), ``data management module'' (\S\ref{sec:data_management}), and ``threat detection module'' (\S\ref{sec:threat_detection}). 

The data collection module is the foundation of the detection systems. It needs to be able to collect system-level provenance information efficiently and accurately. 
The data management module acts as a bridge between the collector and detector. It is responsible for providing efficient and fast query interfaces while storing massive amounts of data efficiently and economically.
The threat detection module needs to process large amounts of data and locate stealth malicious behaviors with the lowest possible overhead and the shortest latency. 

To design such an ideal provenance graph-based detection and investigation system, we should take the following four research questions into consideration:

\textbf{RQ1: }How to reduce the size of the data storage as much as possible while maintaining the semantics?

\textbf{RQ2: }How to balance the space efficiency of the provenance graph storage with the time efficiency of the query?

\textbf{RQ3: }How to design an efficient and robust intrusion detection algorithm and balance the true-positives and false positives?

\textbf{RQ4: }How to shorten the response time of detection and forensics as much as possible? 

The potential answers to RQ1 and RQ2 are discussed in \S\ref{sec:data_management}. The potential answers to RQ3 and RQ4 are discussed in \S\ref{sec:threat_detection}. All in all, this survey makes the following 

\vspace{0.06in}
\noindent \textbf{Contributions}:
\vspace{-\topsep}
\begin{list}{\labelitemi}{\leftmargin=1.5em}
 \setlength{\topmargin}{0pt}
 \setlength{\itemsep}{0em}
 \setlength{\parskip}{0pt}
 \setlength{\parsep}{0pt}
\item We present the first thorough survey for threat detection and investigation with provenance graphs.
\item We taxonomize various representative techniques used in existing papers and depict a typical architecture design of the provenance-based threat detection systems today in \S\ref{subsec:typical-design}.
\item We employ various performance indicators to systematically compare dozens of existing detection systems in \S\ref{sec:threat_detection}. Based on the comparison, we identify the strategy of technology selection for real-world deployment in \S\ref{sec:discussion}. Moreover, we provide multiple insights and challenges for future studies.
\end{list}

The rest of the paper is organized as follows: Section \S\ref{sec:background} introduced the background knowledge of the system-level provenance graph, including several basic definitions, the typical design of a detection system, and brief research history. 
Section \S\ref{sec:relatedworks} introduced related works and the scope of this survey. 
Section \S\ref{sec:data_collection}, \S\ref{sec:data_management}, and \S\ref{sec:threat_detection} focused on three sub-modules, respectively.
Section \S\ref{sec:discussion} detailed described the advantages and disadvantages of different approaches from several perspectives and provided multiple insights and challenges. Section \S\ref{sec:conclusion} presented conclusions.

}

\section{\revision{Background}}
\label{sec:background}

\begin{figure*}
    \centering
    \includegraphics[width=7.1in]{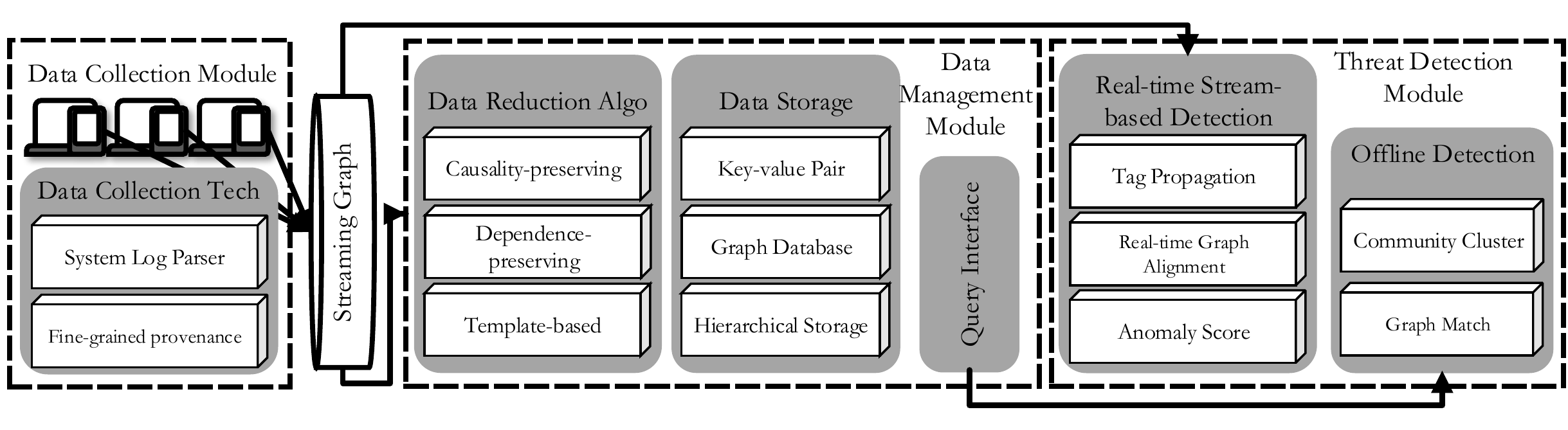}
    \caption{A general framework of provenance-based threat detection system.}
    \label{fig:frameword}
\end{figure*}

\subsection{Definition of System-level Provenance Graph}

System-level provenance graphs treat all system-level entities as vertices and all operations between entities as edges. The operations are collected by auditing tools and generate events stream with timestamps. The order of events affect semantics, and events are directed, which indicate the flow of data or control. Thus, provenance graphs have strong spatial and temporal properties. Such properties are called \textit{causality} for provenance graphs. Correspondingly, provenance graphs are also called \textit{causality graphs}. A series of related basic definitions are given as follows: 

DEFINITION 1. \textbf{Subjects and Objects.} Subject refer to the entity in the system that perform a operation to another entity that is called object. Subject and objects are denoted by $u$ and $v$ respectively.

It is worth mentioning that subjects and objects are relative, a subject of one operation can be the object of another event. Subjects can be processes, threads, etc. Moreover, Objects can be files, sockets, and so on. For different operation systems, the types of subject and object could be different. For example, Windows has unique registry objects and COM objects. However, it is not complicated to extend the provenance graph with more types of subjects and objects.


DEFINITION 2. \textbf{Events} refer to the operations between entities in the system. An event includes four main attributes: the subject performing the operation, the object being operated, the time when the event occurred, and the specific content of the operation. Thus, a event can be denoted by a quad $<subject, object, time, operation>$ (or $<u, v, t, o>$ for short.) Table \ref{tab:events} lists the most commonly used events. And it is relatively easy for analysts to add more events.

DEFINITION 3. \textbf{Provenance Graph} is the collection of all subjects, objects, and events, which can be denoted by $G = (S, O, E)$, where $S$ represent the collection of subjects, $O$ represent the collection of objects, $E$ represent the collection of events. 

In provenance graphs, both subjects and objects are represented as nodes, while events are represented as edges.  There could be more than one edge between two nodes with different time or operation. 

DEFINITION 4. \textbf{Causality Dependency.} Two events $e_1 = (u_1, v_1, t_1)$ and $e_2 = (u_2, v_2, t_2)$ have causality dependency, if $ v_1 = u_2 \wedge t_1 < t_2$.

Causality dependencies indicate the possible data and control flow between two events. However, two events are causality dependent does not necessarily mean there are data or control flow between them. Thus, compared to taint analysis \cite{newsome2005dynamic}, the causality-based analysis will introduce more false dependencies and cause more severe explosion problems.

DEFINITION 5. \textbf{Backward Tracking.} Starting from a single detection point (e.g., a suspicious file), the backward tracking process tries to find all nodes in the provenance graph that causally affect the detection point. 

DEFINITION 6. \textbf{Forward Tracking.} Starting from a single detection point, the forward tracking process tries to find all nodes in the provenance graph that causally depend on the detection point. 

The backward and forward tracking is widely used together in attack investigation to find the entry point and analysis the impact of the attack.

\begin{table}[]
\centering
\caption{Common Provenance Events List}
\vspace{0.03in}
\label{tab:events}
\begin{tabular}{cc}
\toprule Sample graph & Description \\ \midrule
\makecell{\includegraphics[scale=1.2]{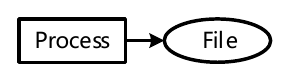}} & Write File \\
\makecell{\includegraphics[scale=1.2]{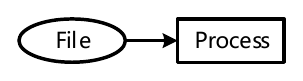}} & Read File \\
\makecell{\includegraphics[scale=1.2]{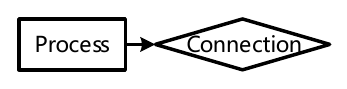}} & Send Data \\
\makecell{\includegraphics[scale=1.2]{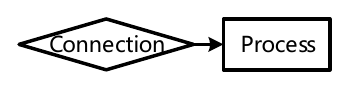}} & Receive Data \\
\makecell{\includegraphics[scale=1.2]{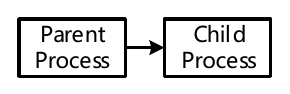}} & Create New Process \\
\makecell{\includegraphics[scale=1.2]{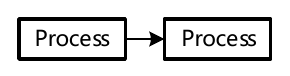}} & \makecell{Inter-process\\communication}\\
\bottomrule 
\end{tabular}
\end{table}

\subsection{Typical Design of Provenance Graph-based Detection System}
\label{subsec:typical-design}

In this section, we will introduce the composition of a typical provenance graph-based detection system. As Figure \ref{fig:frameword} shows, firstly, data collection modules should be installed in the target hosts to collect operations between system objects, which indicates provenance information. Coarse-grained provenance (\S\ref{subsec:coarse}) information can be obtained with built-in auditing systems for most of today's operation systems, such as ETW \cite{ETW} (Event Tracing for Windows) and Linux auditing system \cite{auditd}. However, to collect more fine-grained provenance (\S\ref{subsec:fine}), analyzer needs to install extra infrastructure, such as common libraries or hook into system calls. These fine-grained techniques have much higher overhead, ranging from $2\times$ to $10\times$, and sometimes require support from vendors. The collected information will be parsed into a stream of events defined by Definition 3. The event stream will be transformed into a data management module or directly to a stream-based detection system.

In the data management module, a filter will apply different data reduction algorithm (\S\ref{subsec:data_reduction}) to remove redundant events according to different principles. Data reduction for provenance graph can not only reduce storage space but also reduce subsequent detection or investigation overhead. The compressed data will be stored in databases, which is appropriately designed to support frequent queries (\S\ref{subsec:query}) and persistent access (\S\ref{subsec:storage}). 

The last and most important module is threat detection modules (\S\ref{sec:threat_detection}). Intrusion detection based on provenance graphs is not straight-forward. The most significant challenge comes from the massive amount of data generated in real time. A typical operating system will perform massive file read and write and network connection operations, which brings a lot of background noises. According to the survey results in \cite{xu_high_2016}, for a typical bank with 20,000 hosts, about 70 PB of logs are generated annually. How to find out suspicious events timely is also challenging. One mitigation strategy for both challenges is to build a concise yet comprehensive model incrementally with stream data input.

\subsection{A Brief History of The Adoption of Provenance Graph in Threat Detection}

\begin{figure}
    \centering
    \includegraphics[width=2.6in]{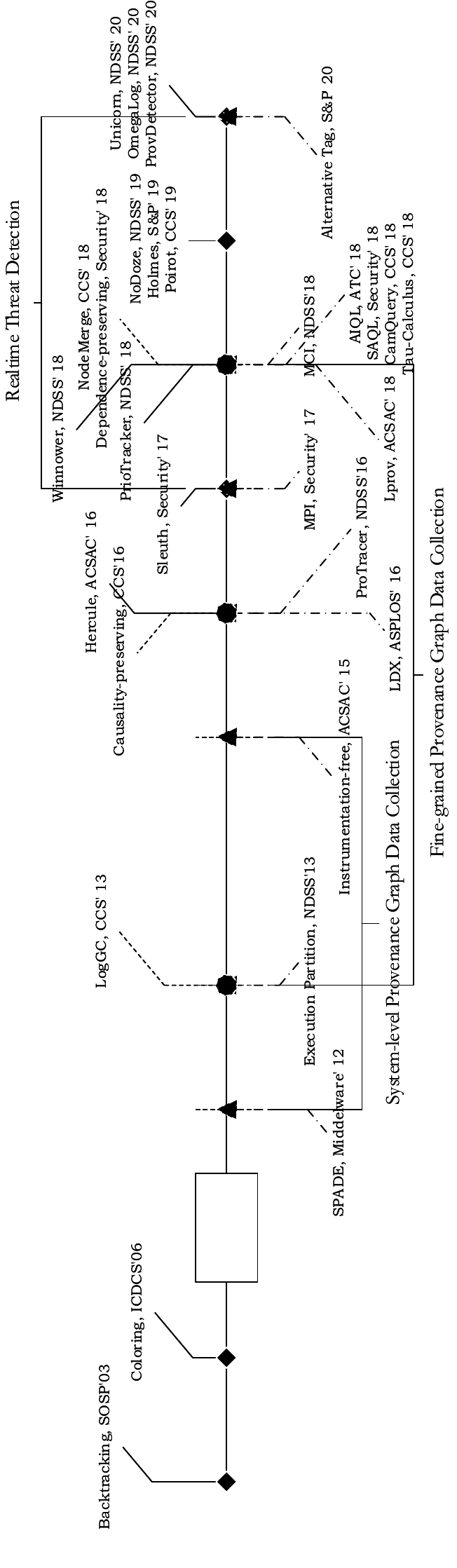}
    \caption{\revision{A brief history of the adoption of provenance graph in threat detection}}
    \label{fig:timeline}
\end{figure}

As shown in Figure \ref{fig:timeline}, we studied dozens of research work and observed two major technology trends. The first trend we find is \textit{the study of fine-grained provenance graph collection}. The original system-level provenance graph is coarse-grained, which has lots of false dependence and thus leads to the ``dependence explosion'' problem. Fine-grained data collection can fundamentally mitigate this problem, while the overhead is much higher. 

The second trend we find is \textit{the study of realtime threat detection}. The response time is critical to real-world security investigation. For example, a quick response can effectively avoid the same attack and reduce the loss. However, the investigations after building the complete provenance graph introduce a long delay to such responses. So far, researchers have been focusing on streaming graph-based detection that can perform real-time detection and investigation.

\revision{
\section{Related Work and Scope of this Survey}
\label{sec:relatedworks}

In this section, we present a holistic view of researches related to system-level provenance graph-based threat detection. Then, we define the scope of this survey and describe our survey methodology.

\subsection{Intrusion Detection}

Intrusion detection \cite{TeresaF.Lunt1993, Buczak2016, Axelsson2000} has been widely studied for several decades on different platforms , such as Host, cloud \cite{Modi}, Mobile platform, Cyber-Physical systems \cite{Mitchell2014}, etc. 

In general, intrusion detection approaches can be divided into three categories: signature-based, anomaly-based, and hybrid. Signature-based approaches \cite{Edge2009} are effective for detecting known attacks without many false-positives. However, lots of labor are required to maintain the signature database. Anomaly-based approaches \cite{Yu2012, Prasad2009, Hodge2004} model the normal behavior and identify anomalies. They can not only detect zero-day attacks but can also produce lots of false-positives. Hybrid approaches combine multiple detection techniques to improve accuracy. 

System-level provenance graph-based detection has a similar taxonomy. However, it utilizes a brand new data source, namely, system-level provenance graphs. Previous low-level data sources, such as system calls and taint analysis, suffer from high overhead and difficulty constructing semantic. In parallel, high-level data sources, such as system audits \cite{auditd}, miss many behaviors and are easily bypassed. The system-level provenance graphs are believed to have the appropriate granularity. It can model all data flow and information flow in systems as graphs containing very rich semantic for intrusion detection. Moreover, it is relatively lightweight to be collected and analyzed in real-time.

Furthermore, attack techniques have evolved, becoming increasingly stealthy and persistent. It is difficult to distinguish malicious behavior based on single-point detection accurately. Correlation analysis \cite{Husak2019, Ficco2013} can combine multi-source information and detect stealth threats effectively without much false-positive. However, most of the existing event-level correlation analyses rely on prior knowledge, therefore, hard to expand. Provenance graph-based detection supports correlation analysis naturally with causality analysis, which can help improve detection accuracy.

\subsection{Provenance}

Data provenance, also called data lineage, was initially introduced to find the origin of data in databases \cite{buneman2001and, woodruff1997supporting}. It provides a historical record of data and its origins. With the provenance information of data, we can obtain the validity and confidence of data.

Data provenances are widely adopted for multiple different purposes, such as reproducibility \cite{greenwood2003provenance, miles2007requirements}, fault injection \cite{naughton2009fault}, and so on. Several surveys have also been done for different provenance applications \cite{simmhan2005survey, freire_provenance_2008, zafar_trustworthy_2017, herschel_survey_nodate}. 

This survey focuses on system-level provenance information modeled as provenance graphs, which record the information flow between system-level objects in detail. Such information can be useful in locating potentially malicious behavior, such as information leakages, etc.

\subsection{Graphs for Security Purpose}

Graph structures are widely utilized in cyber security because of their rich semantics and powerful representation. Different kinds of graphs are extracted for different purpose according to their different properties.

For example, control flow graphs (CFGs) \cite{Venkatasubramanian2003, Petroni2007} and abstract syntax trees (ASTs) \cite{Ndichu2019, Li2019} can effectively model the structure and behavior of programs and are therefore widely used for program analysis and malware detection.
Besides, Bayesian attack graph \cite{Munoz-Gonzalez2016, Wu2012, Frigault2008} can quantify the risks and vulnerabilities in the system to measure the system's security.
Petri net \cite{Xu2006} is a well-known operational model for formal analysis of control and composition of the distributed system. It can formally analyze the security of the system with considerable overhead. 

However, none of the above approaches can effectively model information flow between system-level objects with acceptable overhead, which is critical for system-level threat detection. Therefore, in this paper, we focus on the system-level provenance graph, which can not only track the information flow but also support correlation analysis naturally. It is believed to be the next generation of detection technology.

\subsection{Survey Methodology}

Multiple databases are used to conduct this survey. There are many keywords relevant to our topic, including provenance, causality, audit, logging, detection, forensic, investigation, apt, reduction, collection, etc. However, these keywords are also widely used in other fields. Thus, searching with a single keyword does not work well. As a first step, we searched for several sets of keywords on Google Scholar, including ``provenance + causality + collection'', ``provenance + causality + reduction'', and ``provenance + causality + detection'', which corresponded to three sub-modules in the typical design. 

Nevertheless, the first round of search with keywords combination will miss lots of related works. Thus, we adopted a knowledge graph tool for research papers, namely, Connected Papers \cite{connectedpapers}. This tool is able to search related papers according to not only the citation tree but also the co-citation and bibliographic coupling. It will construct a knowledge graph for every input paper, with this basis, we are able to find lots of related works.
Finally, a number of articles are located with snowball methods. 

}
\section{Data Collection Module}
\label{sec:data_collection}

\begin{table*}[]
\centering
\caption{Comparison of Provenance Data Collection Approaches}
\label{tab:collection}
\begin{tabular}{c|c|c|c|c}
\hline
 & O/H & Acc. & Granularity & \makecell{Other\\Requirements} \\ \hline
\makecell{System-level\\ \cite{gehani_spade:_2012, pasquier_practical_2017}} & Low & Low & Coarse & None \\ \hline
\makecell{Execution Partition\\ \cite{ma_mpi:_2017, lee_high_nodate, ma_accurate_2015, yanguiscope, ma_protracer:_2016}} & Low & Mid & Mid & Instrumentation \\ \hline
\makecell{Causality Inference\\ \cite{kwon_ldx:_2016, kwon_mci_2018, hassan_towards_2018}} & Mid & Mid & Mid & \makecell{Training or \\Dual-Exectuion} \\ \hline
\makecell{Taint Analysis\\ \cite{ji_rain:_2017, kemerlis_libdft:_2012, yang_enabling_2018}} & High & High & Fine & \makecell{Tainting\\Framework} \\ \hline
\makecell{Multi-layer \cite{hassan_omegalog_nodate}} & Low & Low & Coarse & Static Analysis \\ \hline
\end{tabular}
\end{table*}

As the first step, security analyzers need to deploy collectors on target hosts to collect provenance information. Generally, there are two kinds of collectors: The coarse-grained collectors that focus on system-level information flow, such as file reads, inter-process communication, and so on; and fine-grained collectors that involve intra-process information flow tracking. We will comparatively introduce the design and mechanism of these two kinds of collectors in \S\ref{subsec:coarse} and \S\ref{subsec:fine}.

\subsection{Coarse-grained Provenance Collection}
\label{subsec:coarse}

Coarse-grained data collectors only track the provenance between system-level objects, also called system-level collectors. The system-level provenance can be obtained from multiple different sources. Most of today's operating systems have built-in audit system, which can provide necessary information flow among system-level objects. There are also third-party collectors, such as FUSE \cite{fuse}. CamFlow \cite{pasquier_practical_2017} adopts LSM \cite{lsm} and NetFilter \cite{netfilter} to hook kernel objects' security data structure on Linux. Ma et al. proposed \cite{ma_accurate_2015} to obtain system event from windows built-in auditing system ETW \cite{ETW}. SPADE \cite{gehani_spade:_2012} provides multiple collector modules for different systems, for example, hooking system call through Auditd \cite{auditd} on Linux and MacFUSE \cite{OSXFuse} on Mac OS, etc. 

For different operation system and audit tools, the event list could be different. For Linux, all objects are abstracted as files. Table \ref{tab:events} shows the simplest provenance events list. For windows, reading and writing to the registry is important. However, such extension is trivial and will not affect later data management and detection too much. W3C Prov-DM \cite{prov-dm} provide more specific definition. 
In practice, security analyzer should customize the events list to reach a balance between overhead and functionality.

\subsection{Fine-grained Provenance Collection}
\label{subsec:fine}


One common challenge for causality tracking with provenance graph is the "Dependence Explosion" problem, which causes a large number of benign nodes marked as malicious and brings a lot of computing overhead and human labor. Specifically, for a provenance node with $m$ input edges and $n$ output edges, there could be as much as $m\times n$ possible information flows. Fine-grained provenance collectors can solve the "Dependence Explosion" problem fundamentally by associating inputs and outputs more accurately. Ideally, the number of information flow can be reduce to $m+n$. Thus, researchers proposed lots of approaches to collect fine-grained provenance, as shown in Table \ref{tab:collection}. 

Taint analysis that can accurately track information flow within processes are widely used to prevent information leak or zero-day attacks \cite{clause2007dytan, enck2014taintdroid, newsome2005dynamic, xu2006taint}. By combining inter-process provenance analysis and intra-process analysis, researchers \cite{ji_rain:_2017, kemerlis_libdft:_2012, yang_enabling_2018} are able to accurately track the information flow. However, taint analysis introduces significant overhead, slowing down programs by $2\times$ to $10\times$ or more. 

Excessive overhead makes Taint infeasible for large-scale threat detection. To reduce the overhead, Ma et al. \cite{ma_accurate_2015} first proposed execution partition-based approach. They figure out that taint analysis, which tracking information flow between variables, is too fine-grain and not necessary to build causality connection between inputs and outputs. Thus, they try to find a middle ground between coarse-grain processes and fine-grain variables, called \textit{unit}. Many later works \cite{ma_protracer:_2016, ma_mpi:_2017, lee_high_nodate, yanguiscope} adopt a similar idea. All these works make a different assumption about what kind of unit the causality should be maintained in. For example, \cite{ma_accurate_2015} believes that processes can be split into many main loops, and each loop completes a task. Thus, the causality relationship will only be built in the loop. However, such assumptions do not always hold, and these approaches either need extra infrastructure or support from vendors.

Besides improving the accuracy of information flow tracking, causality inference can also effectively reduce false positives. Kwon et al. proposed dual execution-based causality inference \cite{kwon_ldx:_2016, kwon_mci_2018}. By comparing the output buffer contents of the master and slave at the sink(s), they can determine if the sink(s) are causally dependent on the source(s). Hassan et al. proposed Winnower \cite{hassan_towards_2018} that tries to infer the connection by training a model to succinctly summarize the behavior of many nodes. 

After this module, the collected provenance information can be transmit directly to detection module (\S\ref{sec:threat_detection}) or through a data management module (\S\ref{sec:data_management}) first.
\section{Data Management Module}
\label{sec:data_management}

Ubiquitously monitoring system in an organization or enterprise will generate massive amount of data. An ideal data management module should consider how to reduce storage cost while providing effective query interface. In this section, we introduce how to design such an ideal data management module from 3 aspects: data storage models (\S\ref{subsec:storage}), data reduction algorithms (\S\ref{subsec:data_reduction}), and query interface (\S\ref{subsec:query}), and try to answer two research questions, namely, RQ1: How to reduce the size of the data storage as much as possible while maintaining the semantics and RQ2: How to balance the space efficiency of the prove-nance graph storage with the time efficiency of the query?

\begin{figure*}
    \centering
    \includegraphics[width=4.6in]{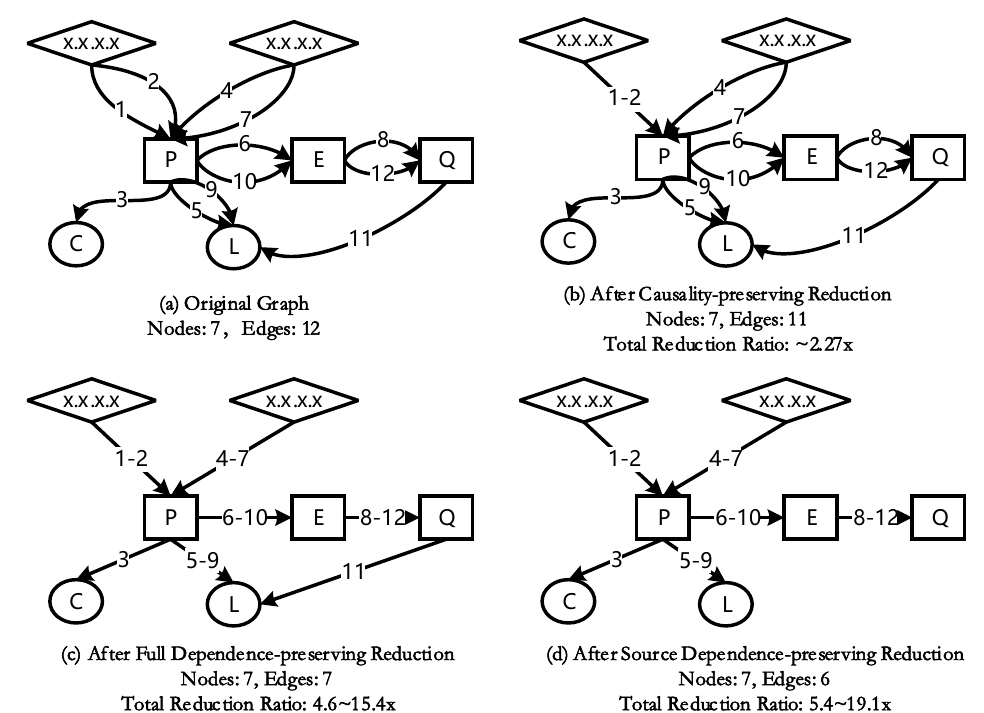}
    \caption{Data reduction algorithms for edges}
    \label{fig:edge_reduction}
\end{figure*}

\subsection{Data Storage Models}
\label{subsec:storage}

The data storage model is the foundation of the whole data management module. The data model used depends on subsequent operations. We will systematically analyze the relationship between different detection algorithms and their corresponding data models in \S\ref{sec:threat_detection}.

A straightforward idea is to store provenance graph with a graph database. Graph database \cite{graph_database} is a widely used NoSQL database, which stores all data as nodes and edges, and provide semantic query interfaces with nodes and edges. Thus, performing graph algorithms, such as backtracking and graph alignment, is relatively easy. However, existing graph database needs to load the whole graph database in the main memory to enable queries. In a large organization, terabytes of data needs to be loaded for a long-running attack campaigns. Even though allocating such large memory is still possible, such approaches incur significant I/O overhead. To mitigate this challenge, the security researchers design detection algorithms \cite{hossain_sleuth:_2017, milajerdi_poirot:_2019, han_unicorn_2020, milajerdi_holmes_2019} that consume every event in the stream only once, and adopt state stored in cache to represent the event history. Corresponding to the cached graph stored in memory, we call the input of such approaches as \textit{streaming graph}.

Vertex-centric database, built on relational database, store all entries as $<K,V>$ pairs, where $K$ is a identifier representing vertexes (nodes) and $V$ is a list of several entries, such as parents nodes, child nodes, and rules \cite{xie_pagoda:_2018}. Such data model can easily count interaction between nodes, thus widely used in abnormal analysis-based detection systems. Furthermore, relational database can be stored in disk and accelerated with in-memory cache, and thus more feasible than graph database-based approaches.

\subsection{Data Reduction Algorithms for Provenance Graphs}
\label{subsec:data_reduction}

In recent years, more organizations, enterprises, and government agencies suffered from advanced persistent threat (APT) attacks [23-25]. These attacks often have multiple phases and last for quite some time. Moreover, these attacks are often very covert and difficult to detect. It has been reported that the average duration of advanced persistent threat attacks lurking within an enterprise is as long as 188 days [26]. However, the amount of data collected in the provenance graph is extremely large, and the amount of data for a single machine can easily exceed 1GB in one day. Moreover, the number of hosts in a large enterprise or organization can reach tens of thousands. This thus brings significant data storage overhead. At the same time, a massive amount of data also brings great difficulties to subsequent data backtracking. Therefore, the algorithm for compressing the provenance graph is a subject that researchers need to study.

The provenance graph is a special graph whose data mainly includes two parts: nodes (subjects and objects) and edges (events). The essence of the compression of the provenance graph is to remove as many unnecessary nodes and edges as possible while maintaining as much semantics as possible. Specifically, three questions need to be considered: 1) How to define the semantics that needs to be maintained? 2) What is the computational complexity of the compression algorithm? 3) How effective is the compression algorithm? With these three questions in mind, we discuss how to compress nodes and edges, respectively.

In this section, we mainly focus on data reduction methods, which refer to some data reduction principle with a guarantee of limited semantic loss.

\subsubsection{Data Reduction for Edges}

In a typical operating system, processes and file objects will exist for a while and generate lots of operations between them. Thus, the number of edges is much larger than that of nodes in most provenance graphs, especially for long-running systems. Data reduction algorithms for edges shall introduce higher data reduction ratios than the algorithms for nodes.

Data reduction approaches need to handle the trade-off between data compression ratio and semantic retention. It is almost impossible to prune data without losing any semantics. Thus, researchers should consider how much semantics should be preserved after data reduction. Causality-preserving reduction approach \cite{xu_high_2016} and dependency-preserving reduction approach \cite{hossain_dependence-preserving_2018} are proposed to define the loss. A simple and intuitive definition of causality is that the first write to an object will affect the subsequent readings.

\textbf{Causality-Preserving Reduction (CPR). } As we discussed in \S\ref{sec:introduction}, causality analysis is the most commonly used operation in provenance graph. Xu et al. proposed causality-preserving reduction \cite{xu_high_2016} that maintains the ability to causality analysis on provenance graphs. A simple and intuitive definition of causality is that the first write to an object will affect the subsequent readings. Thus, to avoid changing the causality between objects, CPR will only remove any repeated writes/reads between a pair of objects with no read/write to the destination object. CPR can completely preserve the topology of the graph, and ensure that most detection algorithms are still valid on the compressed graph. However, the algorithm will lose statistical information, including the access frequency, etc. In real-world scenarios, analyzers should pick reduction algorithms according to the subsequent analysis. 

\textbf{Full Dependence-Preserving Reduction (FDR) and Source Dependence-Preserving Reduction (SDR).} As Figure \ref{fig:edge_reduction} shows, while CPR preserves the semantics in provenance graphs well, it has limited data reduction ratio. To further compress the provenance graph, Hossain et al. \cite{hossain_dependence-preserving_2018} proposed dependence-preserving data compaction.   Dependence-preserving reduction only considers the basic operation on provenance graphs, namely, backward tracking and forward tracking. FDR and SDR rely on global reachability of provenance graphs, which is much more expensive to compute than CPR. To overcome these computational challenges, they proposed versioned dependence graphs, which are widely used to simplify computation produce of provenance \cite{noauthor_provenance-aware_2016, chavan_towards_2015}. 





\begin{figure*}
    \centering
    \includegraphics[width=6.9in]{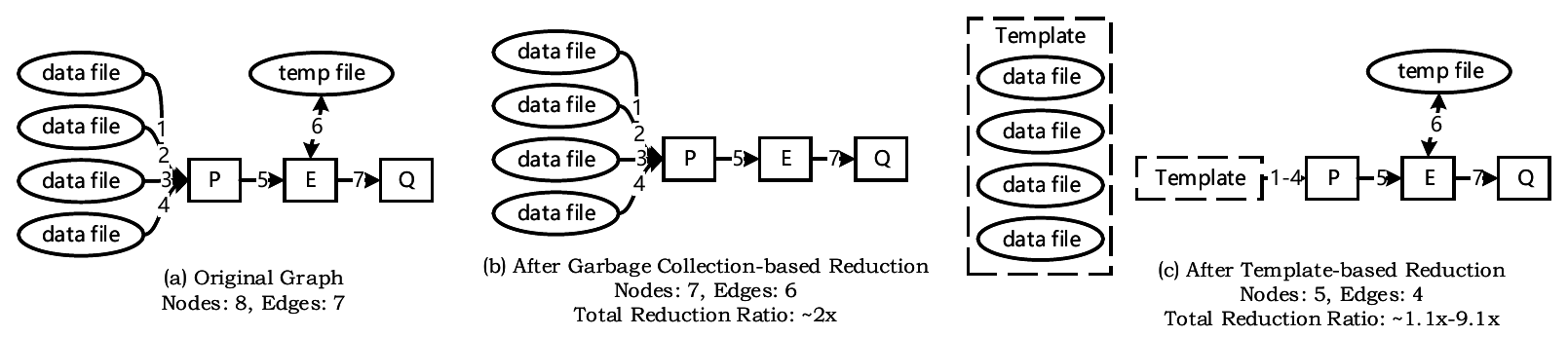}
    \caption{Data reduction algorithms for nodes}
    \label{fig:node_reduction}
\end{figure*}

\subsubsection{Data Reduction for Nodes}

Some techniques try compressing provenance logs via web graph compression algorithm \cite{chapman2008efficient} or detecting common sub-graphs and compressing them \cite{xie2012hybrid}. The main problem of these techniques is that they involve expensive runtime overhead. However, system-level provenance graphs expand quickly. Thus, light-weight compression algorithms are required.

Towards designing efficient compression algorithms, Lee et al. \cite{lee_loggc:_2013} designed garbage collecting for provenance, which can locate isolated temporary nodes. Removing these nodes will not affect causality in provenance graphs. Tang et al. proposed nodemerge \cite{tang_nodemerge:_2018}, which adopt enhanced FP-growth algorithm to find common access patterns during program initialization. The compression ratio of both algorithms is lower than edge-based reduction algorithm.

\subsection{Query Interface}
\label{subsec:query}

Most detection approaches tend to use naive database query interfaces and fixed data structure to ensure universality. However, for customized attack investigation requirement, the naive query interfaces may not be flexible enough. To fill this research gap, researchers proposed series of provenance graph query systems \cite{gao_aiql:_2018, gao_saql:_2018, shu_threat_2018, pasquier_runtime_2018}.

These query systems provide investigation capabilities that naive databases cannot provide or require extra effect. These capabilities are list as following:

\textbf{Causality Tracking. } Provenance graphs have strong spatial and temporal properties, which is thus different from ordinary graphs. Backward and forward tracking should take these properties, which are called causality, into consideration. Such tracking operations are common tasks in forensic for root cause discovery and impact analysis \cite{king_backtracking_2003}. Almost all the query systems regard the causality tracking as their basic function and provide convenient language or interface support \cite{gao_aiql:_2018, shu_threat_2018, pasquier_runtime_2018}. 

\textbf{Provenance Graph Pattern Matching. } Graph pattern matching is at the core of graph query. For threat detection with provenance graph, graph patterns can be used to represent attack behaviors with rich semantics. Thus, pattern matching is equivalent to threat detection. Shu et al. \cite{shu_threat_2018} points out that an ideal pattern matching system should be able to treat patterns as values and compose larger patterns based on others to enable pattern reuse and abstraction. To accomplish such targets, Shu et al. adopt well-designed query language and typing system. 

\textbf{Stream-based Query. } Threat detection is a time-critical mission. To reduce the delay between the attack and the investigation and response, Gao proposed SAQL \cite{gao_saql:_2018}, which is able to take real-time event feed aggregated from multiple hosts as input and provide rich interface. They built the query engine on the top of Siddhi \cite{Siddhi} to leverage its mature stream management engine. To tackle the scalability challenge, they designed a master-dependent query scheme that identiﬁes compatible queries and groups them to use a single copy.

\textbf{Anomaly Analysis. } Security log auditing and threat detection rely heavily on expert experience. In order to adopt domain knowledge from expert to express anomalies, Gao provides a domain-speciﬁc query language, SAQL \cite{gao_saql:_2018}, which allows analysts to express models for (1) rule-based anomalies, (2) time-series anomalies, (3) invariant-based anomalies, and (4) outlier-based anomalies.

All in all, the query systems provide analysts with a thorough attack investigation capability. These systems typically build on mature stream processing system or database, but take provenance graphs' special properties into consideration with specifically-designed data model and query language.  
\section{Threat Detection Module}
\label{sec:threat_detection}

Using the traceability diagram, security analysts can link causal events and entities in the host to obtain a good abstraction ability, which can well describe the data flow and control flow in the system. In order to connect multiple points involved in an attack, the simplest method is to backtrack [22, 30]. However, the simple backtracking algorithm is difficult to distinguish normal data flow from malicious control flow. There is a problem of dependence explosion, so the accuracy is very low. In order to solve this problem and provide a real-time, efficient, and low false positive threat detection system, researchers have proposed many different schemes. In this section, we first give several threat models (\S\ref{subsec:attack}) commonly used in threat detection research using provenance graphs. Then we give a comparison of the existing intrusion detection systems and try to answer two of the research questions we summarize in \S\ref{subsec:typical-design}: RQ3: How to design an efficient and robust intrusion detection algorithm and balance the true-positives and false positives? and RQ4: How to shorten the response time of detection or traceability forensics as much as possible?

\subsection{Attack Models}
\label{subsec:attack}

\subsubsection{Multi-Stage APT Attack (APT) Model}

A large part of threat detection using a provenance graph aims at detecting advanced persistent threat (APT) attacks. APT attacks have the characteristics of advancedness, complexity, concealment, and persistence. Typical APT attacks can be divided into multiple stages, as ATT\&CK Metrics \cite{Mitre} shows. Every stage has a particular target and a variety of different technologies to achieve the target. Real-world attacks usually involve three or more stages. Thus, even if missing some stages, security analyzers can still identify a threat and complete the missing piece with digital forensic techniques. Meanwhile, analyzers can also adopt the multi-stage feature to filter out false alerts.

\subsubsection{Information Leakage (Leakage) Model}

The information leakage model assumes that the attackers are able to take control the entire target system. The goal is to pass the specified sensitive information to endpoints controlled by the attacker in various ways. A large part of APT attacks is also aimed at information leakage. However, unlike the multi-stage APT attack model, the information leakage model does not focus on specific attack technologies, but focuses on the information flow in the system, and continuously monitors whether sensitive information flows to unauthorized points.

\subsubsection{General Attack (General) Model}

General attacks are much more diverse. There are low and stealth attacks like APT but also quick and overt attacks such as ransomware. The target could be stealing information but also pure destruction. Thus, more general and detailed attack models are required to detect such attacks.

\begin{table}[]
\centering
\caption{Taxonomy of Existing Provenance Graph-based Threat Detection System Designs}
\label{tab:threat_detection}
\begin{adjustbox}{angle=90}
\begin{tabular}{lcccccccccc}
\toprule
Approaches & \makecell{Attack\\Models} & Detection Models & Data Models & \makecell{Alert\\Detection} & \makecell{Alert\\Correlration} & \makecell{Response\\Time} & Overhead & \makecell{True\\Positive} & \makecell{False\\Positive} \\
\midrule
Back-tracking \cite{king_backtracking_2003} & General & Naive Backtracking & Cached Graph & \xmark & \cmark & Long & Mid & - & High \\
HERCULE \cite{pei_hercule:_2016} & General & Community Detection & Cached Graph & \xmark & \cmark & Long & Low & - & High \\
POIROT \cite{milajerdi_poirot:_2019} & APT & Graph Alignment & Streaming Graph & \cmark & \cmark & Short & Mid & Mid & Low \\
Log2vec \cite{liu_log2vec:_2019} & General & Graph Embedding & Cached Graph & \cmark & \xmark & Long & Low & Mid & Mid \\
ProvDetector \cite{wang_you_nodate} & General & Graph Embedding & Cached Graph & \cmark & \xmark & Long & Low & Mid & Mid \\
UNICORN \cite{han_unicorn_2020} & APT & Graph Sketch Cluster & Cached Graph & \cmark & \xmark & Mid & Low & High & High \\
PrioTracker \cite{liu_towards_2018} & APT & Anomaly Scores & Cached Graph & \cmark & \xmark & Mid & Low & Mid & Mid \\
NoDoze \cite{hassan_nodoze:_2019} & APT & Anomaly Scores & Vertex-centric DB & \cmark & \xmark & Mid & Low & Mid & Mid \\
P-gaussian \cite{xie_p-gaussian_2019} & APT & Anomaly Scores & Vertex-centric DB & \cmark & \xmark & Mid &  Mid & Mid & Mid\\
Pagoda \cite{xie_pagoda:_2018} & APT & Anomaly Scores & Vertex-centric DB & \cmark & \xmark & Mid & Low & Mid & Mid \\
SWIFT \cite{UlHassan2020} & APT & Anomaly Scores & Vertex-centric DB & \cmark & \xmark & Mid & Low & Mid & Mid \\
Coloring \cite{jiang_provenance-aware_2006} & General & Process Coloring  & Cached Graph & \xmark & \cmark & Long & Low & - & High \\
SLEUTH \cite{hossain_sleuth:_2017} & Leakage & Tag Propagation & Streaming Graph & \cmark & \xmark & Short & Mid & High & High \\
HOLMES \cite{milajerdi_holmes_2019} & APT & Tag Propagation & Streaming Graph & \cmark & \cmark & Short & Low & Mid & Low \\
MORSE \cite{hossaincombating} & APT & Tag Propagation & Streaming Graph & \cmark & \cmark & Short & Low & Mid & Low \\
\bottomrule
\end{tabular}
\end{adjustbox}
\end{table}

\subsection{Threat Detection and Investigation System Design}
\label{tab:detection}

Provenance graphs are able to link events in system with causality, regardless of the time between events, which thus have a overall view of entire attacks. Backtracking, proposed by King \cite{king_backtracking_2003}, is the earliest and most fundamental attack investigation method on provenance graph. Given a detection point, backtracking is able to traverse the whole historical context of system execution. However, naive backtracking requires complete provenance graph and too much human intervention, which thus is neither timely nor efficient.

An ideal threat detection system needs to consider three attributes at the same time: fast response, high efficiency, and high accuracy. However, the size of a provenance graph, even pruned, is very large. Therefore, threat detection on provenance graphs could introduce high space and computing overhead. In order to find a balance between the three attributes, researchers have made many attempts. These approaches can be divided into 3 categories according to the main detection design. 

Firstly, \textit{tag propagation-based approaches} \cite{milajerdi_poirot:_2019, hossain_sleuth:_2017} try to store system execution history incrementally in tags and utilize tag propagation process to trace the causality. These algorithms have roughly linear time complexity. Moreover, they can take streaming graph as input and respond fast. Secondly, \textit{abnormal detection} \cite{liu_towards_2018, hassan_nodoze:_2019, xie_pagoda:_2018, xie_p-gaussian_2019} try to identify abnormal interaction between nodes. Thus, these approaches will model normal behaviors by collecting historical data or data from parallel systems. Finally, \textit{graph matching-based approaches} \cite{liu_log2vec:_2019, milajerdi_poirot:_2019, han_unicorn_2020} try to identify suspicious behavior by matching sub-structure in graphs. However, graph matching is computational complex. Researchers try to extract the graphs' features with graph embedding or graph sketch algorithm or use approximate methods.

As shown in Table \ref{tab:threat_detection}, the target attack models, fundamental detection algorithms, and data management model affect each other and basically determine the design of the detection system. We will compare the system properties according to the ideal system properties introduced in \S\ref{subsec:typical-design}.

\subsubsection{Graph Matching-based Detection}

The graph representation ensures the adversarially robustness of provenance graph-based detection approaches. The connections between nodes indicate the relationship between system entities. Nodes close to each other are more likely to serve the same function. Thus, utilizing community detection algorithm, analysts are able to correlate nodes in the same attack scenarios.
Substructures in a provenance graph can completely describe the malicious behavior. Therefore, it is a very straightforward idea to detect by graph matching. However, graph matching is NP-complete problem \cite{de2008subgraph}. Thus, researchers have proposed many approximate methods. 

Milajerdi et al. proposed POIROT \cite{milajerdi_poirot:_2019} and the key online graph alignment algorithm. Utilizing query graph manually extracted from threat intelligence and the graph alignment algorithm, they could locate threats in provenance graph quickly. However, extracting query graphs requires a lot of manual work. Thus, it is difficult to cover all kind of advanced attacks in various forms.

Graph embedding are widely used to extract graph features into vertexs while maximally preserving properties like graph structure and information \cite{yan2006graph, wang2014knowledge, goyal2018graph}. Utilizing the graph embedding, researchers can effectively and efficiently detect threats by separating malicious and benign log entries into different clusters and identifying malicious ones \cite{liu_log2vec:_2019, wang_you_nodate}. 
However, such methods typical work on cached graph, so the response is slower; meanwhile, it requires a lot of training data, so it is not suitable for advanced attacks.

To tackle the above two challenges, Han et al. proposed UNICORN \cite{han_unicorn_2020}, which adopts a historical graph sketch approach to build an incrementally updatable, ﬁxed size, longitudinal graph data structure. So, they can find threats when the graph structure changed.
However, this is an anomaly detection-based approaches, which thus suffers from the limitation of anomaly detection.

\subsubsection{Anomaly Score-based Detection}

Anomaly score-based detection tries to quantify the suspiciousness of each edge between node pairs. Using historical statistics, researchers can find abnormal access in system. Specifically, Pagoda \cite{xie_pagoda:_2018} takes into account the anomaly degree of both a single provenance path and the whole provenance graph. Their subsequence work P-Gaussian \cite{xie_p-gaussian_2019} can detect variants using gaussian distribution scheme. PrioTracker \cite{liu_towards_2018} and NoDoze \cite{hassan_nodoze:_2019} adjust the events' suspiciousness based on its nighber's suspiciousness. 

Compared with graph-based anomaly detection, anomaly score-based detection has much less parameters to tune, which thus is much easier to implement and deploy. Meanwhile, anomaly score-based detection typically adopts a vertex-centric relational database, which is much faster than graph database. 

\subsubsection{Tag Propagation-based Detection}

Tag propagation-based detection can be divided into two phases, namely, tag initialization and tag propagation. In tag initialization phase, tags are assigned to nodes. The amount of nodes is much less than edges. Thus, storing and updating tags is efficient. In tag propagation phase, tags are passed along the edge according to the pre-designed rules. In this phase, different tags could meet at the same node and triage future calculations together.  

Process coloring proposed by Jiang et al. \cite{jiang_provenance-aware_2006} is a simplified tag-based approach. In the tag initialization phase, tags (colors) are assigned to each remotely-accessible server or process. Then, in the tag propagation phase, tags can be inherited by spawned child processes or diffused indirectly through process actions. As a result, analysts can quickly identify the break-in point without tedious backtracking. 

Follow-up works adopt more complex tag design to implement more functions. SLEUTH \cite{hossain_sleuth:_2017} utilizes two types of tags, namely, trustworthiness tags (t-tags) and conﬁdentiality tags (c-tags), to implement a policy enforcement framework. In short, an alarm is triggered when a node with low trustworthiness accesses a node with high confidentiality. Specifically, in the tag initialization phase, t-tags and c-tags are assigned to the nodes according to the predefined trustworthiness and confidentiality respectively. In the tag propagation phase, the trustworthiness and confidentiality are propagated, and the accesses that violate the policy will be captured. 

However, tag propagation-based approaches also suffer from the "dependency explosion" problem. Without extra control, single tag can spread to everywhere and cause a lot of false positives. To tackle this challenge, Milajerdi et al. proposed HOLMES \cite{milajerdi_holmes_2019}, which raises the detection threshold by requiring the aggregation of more tags. In the tag initialization phase, HOLMES assigns fewer tags only to process with suspicious behaviors. These suspicious behaviors contain lots of false positives. Thus, in the tag propagation phase, HOLMES requires multiple tags to aggregate and reach a pre-defined threshold, and then triage the alert. 
Another way to avoid the dependency explosion problem is to make the impact decrease as the number of transmission rounds increases. MORSE \cite{hossaincombating} achieves this with tag decay and tag attenuation techniques.

All in all, tag propagation-based approaches have the following advantages. Firstly, tag initialization and propagation processes replace computationally expensive graph matching algorithm and lower the overhead. Secondly, tag propagation-based approaches take one event at a time and update states correspondingly, which thus can support streaming graph input naturally and can respond quickly. Last but not least, the information stored in tags can be used to locate the point involved in the intrusion quickly, and thus avoid the tedious backtracking algorithm.

\revision{

}


\revision{
\section{Discussion}
\label{sec:discussion}

In this section, we discuss provenance-based threat detection as a whole from several essential perspectives. First, we compare the effects of different combinations of the data and detection model on the performance metrics. Then, we will describe the prevailing dependence explosion problem and possible solutions. Finally, we discuss how different approaches strike a balance between true-positive and false-positive. Moreover, as summarized in Table \ref{tab:insights}, multiple insights and challenges will be provided for real-world practice and future studies.

\begin{table*}[]
\small
\centering
\caption{\revision{Insights and Challenges on Provenance Graph-based Threat Detection}}
\label{tab:insights}
\begin{tabular}{cp{15cm}}
\toprule
Insight 1 & There are common combinations between the detection model and the data model, which enable optimal performance.\\
Insight 2 & The graph database has poor performance. Thus, the cached graph is not recommended in practice.\\
Insight 3 & Tag propagation-based detection can handle the provenance graph as a stream in real-time and thus have the shortest response time.\\
Insight 4 & The dependence explosion problem can be addressed fundamentally by adopting fine-grained data collection methods. \\
Challenge 1 & Existing fine-grained data collection methods involve significant overhead. How to build a low overhead fine-grained collector is still a pressing research problem.\\
Insight 5 & Existing approaches can only mitigate the dependence explosion problem and may involve potential vulnerabilities.\\
Challenge 2 & More efficient and robust algorithm-based solutions are still direly needed for the dependence explosion problem.\\
Insight 6 & For provenance graph-based detection, most existing detection models are sequence-based rather than graph-based.\\
Challenge 3 & Existing sequence-based real-time detection approaches may not be robust enough to distinguish malicious behavior from benign ones accurately. Therefore, it is still necessary to design and implement more robust detection models.\\
Challenge 4 & There is a lacking of unified datasets and data format for provenance graph-based detection.\\
Challenge 5 & There is a lacking of study on potential evasion for provenance graph-based detection.\\
\bottomrule
\end{tabular}
\end{table*}

\subsection{How the Selection of Data Models and Detection Models Will Affect Performance?}

The detection model and the data model are two important parts of the threat detection system, directly determining the performance.  As shown in Table \ref{tab:threat_detection} \textbf{(Insight 1) there are three frequent combinations of two models, namely, ``anomaly score + vertex-centric DB,'' ``tag propagation + Streaming graph,'' and ``cached graph + others.''}

Caching the data as graphs in the graph database is the most intuitive and convenient way. Almost all detection models work on cached graphs. However, \textbf{(Insight 2) it is an inefficient way because of the poor performance of the graph database. Thus, the cached graph is not recommended in practice. }
NoDoze et al. \cite{hassan_nodoze:_2019, xie_p-gaussian_2019, xie_pagoda:_2018} proposed vertex-centric DB, which is essentially a relational database, as an alternative. While adopting vertex-centric makes access to nodes faster, it also makes access to edges more complicated and slower. Thus, NoDoze et al. adopted anomaly scores-based detection approaches that only need to access nodes' information.
SLEUTH et al. \cite{hossain_sleuth:_2017, milajerdi_holmes_2019, milajerdi_poirot:_2019} choose not to cache the provenance graph. Instead, they process all nodes and edges once, and cache processing results in tags. Moreover, they embed the graph structure information in the tag propagation process. By this means, \textbf{(Insight 3) tag propagation-based detection can handle the provenance graph as a stream in real-time and thus have the shortest response time.}

\subsection{How to Solve the Dependence Explosion Problem?}

As discussed in \S\ref{sec:threat_detection}, dependence explosion is a common problem caused by the coarse-grained provenance graph and causality analysis, which will bring extra false-positive and overhead. Hence, \textbf{(Insight 4) we can address the dependence explosion problem fundamentally by adopting fine-grained data collection methods, as discussed in \S\ref{subsec:fine}.} \textbf{(Challenge 1) However, these methods involve significant runtime or development overhead and, therefore, hard to be utilized in real-world scenarios.}

To mitigating the dependence explosion problem, several algorithm-based approaches are proposed based on different assumptions. Nodoze et al. \cite{hassan_nodoze:_2019, xie_pagoda:_2018} assign anomaly scores to each edge based on the frequency with which related events have happened before. Then, the anomaly score will be propagated along the paths. And paths with low anomaly scores will be ignored. 
Their underlying assumption is that an attack will always involve unusual edges in provenance graphs. However, such an assumption not always holds. One real-world example is the gitpwnd attack \cite{gitpwnd}, which completes the attack exclusively with the git workflow. Attackers can intentionally avoid unusual dependencies that trigger such detection.

SLEUTH and subsequent works \cite{hossain_sleuth:_2017, milajerdi_holmes_2019, hossaincombating} adopt tag decay based approaches. These works try to limit the spread of tags by limiting the number of rounds or time of tag propagation. Their underlying assumption is that the attack will perform the attack as soon as possible. Nevertheless, apparently, attackers can bypass such detection by maintaining stealth for a long time or involving more intermediate nodes to extend the attack chain.

All in all, \textbf{(Insight 5) existing algorithm-based approaches can only mitigate the dependence explosion problem and may involve potential vulnerabilities.} Therefore, \textbf{(Challenge 2) more efficient and robust algorithm-based solutions are still direly needed for the dependence explosion problem.} One possible solution is to determine the underlying information flow with high-level semantic information, as discussed in \S\ref{subsec:fine}.

\subsection{How to Balance the True-positive and False-positive?}

True-positive and false-positive are the most critical and fundamental indicators for a detection system. In general, complicated detection models are better at distinguishing malicious and benign behavior and thus having higher accuracy. For example, the multi-stage model utilized by provenance graph-based detection systems can improve accuracy by alert correlation and perform better than the single-point detection model. However, more complicated models tend to have higher overheads as well.

Specifically, \textbf{(Insight 6) for provenance graph-based detection, most existing detection models are sequence-based rather than graph-based}, including tag propagation-based approaches \cite{hossain_sleuth:_2017, milajerdi_holmes_2019, hossaincombating, milajerdi_poirot:_2019} and most anomaly detection approaches \cite{hassan_nodoze:_2019, xie_pagoda:_2018}. While graph-based detection approaches \cite{han_unicorn_2020, pei_hercule:_2016, wang_you_nodate, liu_log2vec:_2019} typically have longer response times and higher overhead.

High true-positive and low false-positive are often contradictory when adopting the same detection model. However, security analyzers can still seek a balance between them through a combination of techniques and parameter tuning. For example, HOLMES \cite{milajerdi_holmes_2019} utilizes relatively simple signatures to cover as many malicious behaviors as possible. Meanwhile, it adopts the alert correlation to filter false alarms. This process involves lots of parameters, whose tuning significantly affects the accuracy and efficiency of systems. Nevertheless, these parameters are often determined empirically, which makes the detection results not stable. 
In comparison, POIROT \cite{milajerdi_poirot:_2019} uses more sophisticated signatures to avoid false-positive. To improve the coverage, they need to collect a signature database from massive real-world threat intelligence. However, such an approach still cannot detect previously unseen malicious behaviors. 

\textbf{(Challenge 3) Existing sequence-based real-time detection approaches may not be robust enough to distinguish malicious behavior from benign ones accurately. Therefore, it is still necessary to design and implement more robust detection models.}

\subsection{Other Practical Challenges}

\textbf{Challenge 4: Lacking of unified datasets and data format.} Unified datasets and data format can significantly lower the barriers for further research, reproduction, and quantitative comparison. However, as far as we know, the only publicly available dataset for provenance graph-based detection is the Engagement 3 and 5 datasets from the Transparent Computing program \cite{transparent-computing}. Most existing work has to rely on limited self-collected attack data. These datasets only contain dozens of attacks, which can hardly represent various sophisticated attacks in the real-world. Thus, it is claimed that a unified dataset and data format are direly needed.

\textbf{Challenge 5: Lacking of study on potential evasion.} Anti-evasion is a core competency for detection systems. Research into the potential evasion problem is essential for new detection mechanisms, making the detection result more reliable. However, such studies are still missing for system-level provenance graph-based detection. 


\subsection{An Ideal Detection Approach}
 
Synthesizing the above discussion, we propose what an ideal system should like:

\begin{list}{\labelitemi}{\leftmargin=1.5em}
 \setlength{\topmargin}{0pt}
 \setlength{\itemsep}{0em}
 \setlength{\parskip}{0pt}
 \setlength{\parsep}{0pt}
 
\item \textbf{A real-time approach:} An ideal detection system should have the lowest possible overhead and the shortest possible response time. Therefore, the system must be able to process streaming provenance graphs without caching too much data. From a performance perspective, tag propagation-based approaches are the best.
\item \textbf{A robust and effective approach:} For a detection system, robustness and effectiveness means that it needs to distinguish malicious behavior from benign ones accurately in any case. That is to say, the detection model should be complicated enough to demonstrate the difference between malicious behavior from benign ones. From this point of view, the graph-based modeling approach is better than the others.
\end{list}

Specifically, we can try to design and implement such a system by answering the research questions mentioned in \S\ref{subsec:typical-design} and the following the pace of existing work detailed described in \S\ref{sec:data_management} and \S\ref{sec:threat_detection}. 

}

\section{Conclusion}
\label{sec:conclusion}

As a system behavior abstraction tool, provenance graphs are widely accepted for endpoint threat detection. In this paper, we present the typical system architecture for provenance graph-based threat detection. Then, we systematically introduced and compared techniques choice involved and concluded existing research challenges for future study. 

\section*{Acknowledgment}
We would like to thank the anonymous reviewers for providing valuable feedback on our work. This work is supported by the Joint Funds of the National Natural Science Foundation of China (U1936215) and the Key Research and Development Program of Zhejiang Province (2018C01088).


\bibliographystyle{elsarticle-num-names} 
\bibliography{Provenance}

\begin{thebibliography}{94}
\expandafter\ifx\csname natexlab\endcsname\relax\def\natexlab#1{#1}\fi
\providecommand{\url}[1]{\texttt{#1}}
\providecommand{\href}[2]{#2}
\providecommand{\path}[1]{#1}
\providecommand{\DOIprefix}{doi:}
\providecommand{\ArXivprefix}{arXiv:}
\providecommand{\URLprefix}{URL: }
\providecommand{\Pubmedprefix}{pmid:}
\providecommand{\doi}[1]{\href{http://dx.doi.org/#1}{\path{#1}}}
\providecommand{\Pubmed}[1]{\href{pmid:#1}{\path{#1}}}
\providecommand{\bibinfo}[2]{#2}
\ifx\xfnm\relax \def\xfnm[#1]{\unskip,\space#1}\fi
\bibitem[{Sym(2020)}]{Symantec}
\bibinfo{title}{Symantec internet security threat report},
  \bibinfo{year}{2020}. \URLprefix
  \url{https://docs.broadcom.com/doc/istr\-24\-2019\-en}.
\bibitem[{tra(0152)}]{transparent-computing}
\bibinfo{title}{Darpa transparentcomputing}, \bibinfo{year}{2015.2}. \URLprefix
  \url{https://www.darpa.mil/program/transparent-computing}.
\bibitem[{Hossain et~al.(2017)Hossain, Milajerdi, Wang, Eshete, Gjomemo, Sekar,
  Stoller, and Venkatakrishnan}]{hossain_sleuth:_2017}
\bibinfo{author}{M.~N. Hossain}, \bibinfo{author}{S.~M. Milajerdi},
  \bibinfo{author}{J.~Wang}, \bibinfo{author}{B.~Eshete},
  \bibinfo{author}{R.~Gjomemo}, \bibinfo{author}{R.~Sekar},
  \bibinfo{author}{S.~Stoller}, \bibinfo{author}{V.~N. Venkatakrishnan},
\newblock \bibinfo{title}{{SLEUTH}: {Real}-time {Attack} {Scenario}
  {Reconstruction} from {COTS} {Audit} {Data}},
\newblock in: \bibinfo{booktitle}{26th {USENIX} {Security} {Symposium},
  {USENIX} {Security} 2017, {Vancouver}, {BC}, {Canada}, {August} 16-18,
  2017.}, \bibinfo{year}{2017}, pp. \bibinfo{pages}{487--504}.
\bibitem[{Milajerdi et~al.(2019)Milajerdi, Gjomemo, Eshete, Sekar, and
  Venkatakrishnan}]{milajerdi_holmes_2019}
\bibinfo{author}{S.~M. Milajerdi}, \bibinfo{author}{R.~Gjomemo},
  \bibinfo{author}{B.~Eshete}, \bibinfo{author}{R.~Sekar},
  \bibinfo{author}{V.~Venkatakrishnan},
\newblock \bibinfo{title}{{HOLMES}: {Real}-{Time} {APT} {Detection} through
  {Correlation} of {Suspicious} {Information} {Flows}},
\newblock in: \bibinfo{booktitle}{2019 {IEEE} {Symposium} on {Security} and
  {Privacy} ({SP})}, \bibinfo{year}{2019}, pp. \bibinfo{pages}{1137--1152}.
  \DOIprefix\doi{10.1109/SP.2019.00026}, \bibinfo{note}{iSSN: 1081-6011}.
\bibitem[{Xie et~al.(2018)Xie, Feng, Hu, Li, Sample, and
  Long}]{xie_pagoda:_2018}
\bibinfo{author}{Y.~Xie}, \bibinfo{author}{D.~Feng}, \bibinfo{author}{Y.~Hu},
  \bibinfo{author}{Y.~Li}, \bibinfo{author}{S.~Sample},
  \bibinfo{author}{D.~Long},
\newblock \bibinfo{title}{Pagoda: {A} {Hybrid} {Approach} to {Enable}
  {Efficient} {Real}-time {Provenance} {Based} {Intrusion} {Detection} in {Big}
  {Data} {Environments}},
\newblock \bibinfo{journal}{IEEE Transactions on Dependable and Secure
  Computing}  (\bibinfo{year}{2018}) \bibinfo{pages}{1--1}.
  \DOIprefix\doi{10.1109/TDSC.2018.2867595}.
\bibitem[{Milajerdi et~al.(2019)Milajerdi, Eshete, Gjomemo, and
  Venkatakrishnan}]{milajerdi_poirot:_2019}
\bibinfo{author}{S.~M. Milajerdi}, \bibinfo{author}{B.~Eshete},
  \bibinfo{author}{R.~Gjomemo}, \bibinfo{author}{V.~Venkatakrishnan},
\newblock \bibinfo{title}{{POIROT}: {Aligning} {Attack} {Behavior} with
  {Kernel} {Audit} {Records} for {Cyber} {Threat} {Hunting}},
\newblock in: \bibinfo{booktitle}{Proceedings of the 2019 {ACM} {SIGSAC}
  {Conference} on {Computer} and {Communications} {Security}}, {CCS} '19,
  \bibinfo{publisher}{ACM}, \bibinfo{address}{New York, NY, USA},
  \bibinfo{year}{2019}, pp. \bibinfo{pages}{1795--1812}. \URLprefix
  \url{http://doi.acm.org/10.1145/3319535.3363217}.
  \DOIprefix\doi{10.1145/3319535.3363217}, \bibinfo{note}{event-place: London,
  United Kingdom}.
\bibitem[{Hassan et~al.(2019)Hassan, Guo, Li, Li, Chen, Jee, Li, and
  Bates}]{hassan_nodoze:_2019}
\bibinfo{author}{W.~U. Hassan}, \bibinfo{author}{S.~Guo},
  \bibinfo{author}{D.~Li}, \bibinfo{author}{Z.~Li}, \bibinfo{author}{Z.~Chen},
  \bibinfo{author}{K.~Jee}, \bibinfo{author}{Z.~Li},
  \bibinfo{author}{A.~Bates},
\newblock \bibinfo{title}{{NoDoze}: {Combatting} {Threat} {Alert} {Fatigue}
  with {Automated} {Provenance} {Triage}},
\newblock in: \bibinfo{booktitle}{{NDSS}}, \bibinfo{year}{2019}. \URLprefix
  \url{https://www.zhichunli.org/publication/NDSS19-NoDoze.pdf}.
\bibitem[{Xie et~al.(2019)Xie, Wu, Feng, and Long}]{xie_p-gaussian_2019}
\bibinfo{author}{Y.~Xie}, \bibinfo{author}{Y.~Wu}, \bibinfo{author}{D.~Feng},
  \bibinfo{author}{D.~Long},
\newblock \bibinfo{title}{P-{Gaussian}: {Provenance}-{Based} {Gaussian}
  {Distribution} for {Detecting} {Intrusion} {Behavior} {Variants} {Using}
  {High} {Efficient} and {Real} {Time} {Memory} {Databases}},
\newblock \bibinfo{journal}{IEEE Transactions on Dependable and Secure
  Computing}  (\bibinfo{year}{2019}) \bibinfo{pages}{1--1}.
  \DOIprefix\doi{10.1109/TDSC.2019.2960353}.
\bibitem[{Ma et~al.(2015)Ma, Lee, Kim, Rhee, Zhang, and Xu}]{ma_accurate_2015}
\bibinfo{author}{S.~Ma}, \bibinfo{author}{K.~H. Lee}, \bibinfo{author}{C.~H.
  Kim}, \bibinfo{author}{J.~Rhee}, \bibinfo{author}{X.~Zhang},
  \bibinfo{author}{D.~Xu},
\newblock \bibinfo{title}{Accurate, {Low} {Cost} and {Instrumentation}-{Free}
  {Security} {Audit} {Logging} for {Windows}},
\newblock in: \bibinfo{booktitle}{Proceedings of the 31st {Annual} {Computer}
  {Security} {Applications} {Conference}}, {ACSAC} 2015,
  \bibinfo{publisher}{ACM}, \bibinfo{address}{New York, NY, USA},
  \bibinfo{year}{2015}, pp. \bibinfo{pages}{401--410}. \URLprefix
  \url{http://doi.acm.org/10.1145/2818000.2818039}.
  \DOIprefix\doi{10.1145/2818000.2818039}.
\bibitem[{Barre et~al.(2019)Barre, Gehani, and Yegneswaran}]{barre2019mining}
\bibinfo{author}{M.~Barre}, \bibinfo{author}{A.~Gehani},
  \bibinfo{author}{V.~Yegneswaran},
\newblock \bibinfo{title}{Mining data provenance to detect advanced persistent
  threats},
\newblock in: \bibinfo{booktitle}{11th International Workshop on Theory and
  Practice of Provenance (TaPP 2019)}, \bibinfo{year}{2019}.
\bibitem[{aud(0203)}]{auditd}
\bibinfo{title}{Linux auditd}, \bibinfo{year}{2020.3}. \URLprefix
  \url{https://linux.die.net/man/8/auditd}.
\bibitem[{Gehani and Tariq(2012)}]{gehani_spade:_2012}
\bibinfo{author}{A.~Gehani}, \bibinfo{author}{D.~Tariq},
\newblock \bibinfo{title}{{SPADE}: support for provenance auditing in
  distributed environments},
\newblock \bibinfo{publisher}{Springer-Verlag New York, Inc.},
  \bibinfo{year}{2012}, pp. \bibinfo{pages}{101--120}.
\bibitem[{Han et~al.(2018)Han, Pasquier, and Seltzer}]{han2018provenance}
\bibinfo{author}{X.~Han}, \bibinfo{author}{T.~Pasquier},
  \bibinfo{author}{M.~Seltzer},
\newblock \bibinfo{title}{Provenance-based intrusion detection: opportunities
  and challenges},
\newblock in: \bibinfo{booktitle}{10th $\{$USENIX$\}$ Workshop on the Theory
  and Practice of Provenance (TaPP 2018)}, \bibinfo{year}{2018}.
\bibitem[{APT(2020)}]{APT}
\bibinfo{title}{Advanced persistent threat}, \bibinfo{year}{2020}. \URLprefix
  \url{https://en.wikipedia.org/wiki/Advanced\_persistent\_threat}.
\bibitem[{Newsome and Song(2005)}]{newsome2005dynamic}
\bibinfo{author}{J.~Newsome}, \bibinfo{author}{D.~X. Song},
\newblock \bibinfo{title}{Dynamic taint analysis for automatic detection,
  analysis, and signaturegeneration of exploits on commodity software.},
\newblock in: \bibinfo{booktitle}{NDSS}, volume~\bibinfo{volume}{5},
  \bibinfo{organization}{Citeseer}, \bibinfo{year}{2005}, pp.
  \bibinfo{pages}{3--4}.
\bibitem[{ETW(0203)}]{ETW}
\bibinfo{title}{Event tracing for windows}, \bibinfo{year}{2020.3}. \URLprefix
  \url{https://docs.microsoft.com/en-us/windows/win32/etw/about\-event\-tracing}.
\bibitem[{Xu et~al.(2016)Xu, Wu, Li, Jee, Rhee, Xiao, Xu, Wang, and
  Jiang}]{xu_high_2016}
\bibinfo{author}{Z.~Xu}, \bibinfo{author}{Z.~Wu}, \bibinfo{author}{Z.~Li},
  \bibinfo{author}{K.~Jee}, \bibinfo{author}{J.~Rhee},
  \bibinfo{author}{X.~Xiao}, \bibinfo{author}{F.~Xu},
  \bibinfo{author}{H.~Wang}, \bibinfo{author}{G.~Jiang},
\newblock \bibinfo{title}{High {Fidelity} {Data} {Reduction} for {Big} {Data}
  {Security} {Dependency} {Analyses}},
\newblock in: \bibinfo{booktitle}{Proceedings of the 2016 {ACM} {SIGSAC}
  {Conference} on {Computer} and {Communications} {Security}}, {CCS} '16,
  \bibinfo{publisher}{ACM}, \bibinfo{address}{New York, NY, USA},
  \bibinfo{year}{2016}, pp. \bibinfo{pages}{504--516}. \URLprefix
  \url{http://doi.acm.org/10.1145/2976749.2978378}.
  \DOIprefix\doi{10.1145/2976749.2978378}.
\bibitem[{{Teresa F. Lunt}(1993)}]{TeresaF.Lunt1993}
\bibinfo{author}{{Teresa F. Lunt}},
\newblock \bibinfo{title}{{A survey of intrusion detection techniques}},
\newblock \bibinfo{journal}{Computer {\&} Security}  (\bibinfo{year}{1993})
  \bibinfo{pages}{405--518}.
\bibitem[{Buczak and Guven(2016)}]{Buczak2016}
\bibinfo{author}{A.~L. Buczak}, \bibinfo{author}{E.~Guven},
\newblock \bibinfo{title}{{A Survey of Data Mining and Machine Learning Methods
  for Cyber Security Intrusion Detection}},
\newblock \bibinfo{journal}{IEEE Communications Surveys and Tutorials}
  \bibinfo{volume}{18} (\bibinfo{year}{2016}) \bibinfo{pages}{1153--1176}.
  \DOIprefix\doi{10.1109/COMST.2015.2494502}.
\bibitem[{Axelsson(2000)}]{Axelsson2000}
\bibinfo{author}{S.~Axelsson}, \bibinfo{title}{{Intrusion Detection Systems: A
  Survey and Taxonomy}}, \bibinfo{type}{Technical Report},
  \bibinfo{year}{2000}.
\bibitem[{Modi et~al.(????)Modi, Patel, Borisaniya, Patel, and
  Rajarajan}]{Modi}
\bibinfo{author}{C.~Modi}, \bibinfo{author}{D.~Patel},
  \bibinfo{author}{B.~Borisaniya}, \bibinfo{author}{A.~Patel},
  \bibinfo{author}{M.~Rajarajan},
\newblock \bibinfo{title}{{A survey of intrusion detection techniques in
  Cloud}},
\newblock \bibinfo{journal}{Journal of Network and Computer Applications}
  \bibinfo{volume}{36} (????) \bibinfo{pages}{42--57}.
  \DOIprefix\doi{10.1016/j.jnca.2012.05.003}.
\bibitem[{Mitchell and Chen(2014)}]{Mitchell2014}
\bibinfo{author}{R.~Mitchell}, \bibinfo{author}{I.~R. Chen}, \bibinfo{title}{{A
  survey of intrusion detection techniques for cyber-physical systems}},
  \bibinfo{year}{2014}. \URLprefix
  \url{https://dl.acm.org/doi/10.1145/2542049}.
  \DOIprefix\doi{10.1145/2542049}.
\bibitem[{Edge and {Falcone Sampaio}(2009)}]{Edge2009}
\bibinfo{author}{M.~E. Edge}, \bibinfo{author}{P.~R. {Falcone Sampaio}},
\newblock \bibinfo{title}{{A survey of signature based methods for financial
  fraud detection}},
\newblock \bibinfo{journal}{Computers and Security} \bibinfo{volume}{28}
  (\bibinfo{year}{2009}) \bibinfo{pages}{381--394}.
  \DOIprefix\doi{10.1016/j.cose.2009.02.001}.
\bibitem[{Yu(2012)}]{Yu2012}
\bibinfo{author}{Y.~Yu},
\newblock \bibinfo{title}{{A SURVEY OF ANOMALY INTRUSION DETECTION TECHNIQUES
  *}},
\newblock \bibinfo{journal}{Journal of Computing Sciences in Colleges}
  (\bibinfo{year}{2012}). \DOIprefix\doi{10.5555/2379703}.
\bibitem[{Prasad et~al.(2009)Prasad, Almanza-Garcia, and Lu}]{Prasad2009}
\bibinfo{author}{N.~R. Prasad}, \bibinfo{author}{S.~Almanza-Garcia},
  \bibinfo{author}{T.~T. Lu},
\newblock \bibinfo{title}{{Anomaly detection}},
\newblock \bibinfo{journal}{Computers, Materials and Continua}
  \bibinfo{volume}{14} (\bibinfo{year}{2009}) \bibinfo{pages}{1--22}.
  \URLprefix \url{http://doi.acm.org/10.1145/1541880.1541882}.
  \DOIprefix\doi{10.1145/1541880.1541882}.
\bibitem[{Hodge and Austin(2004)}]{Hodge2004}
\bibinfo{author}{V.~J. Hodge}, \bibinfo{author}{J.~Austin}, \bibinfo{title}{{A
  survey of outlier detection methodologies}}, \bibinfo{year}{2004}.
  \DOIprefix\doi{10.1023/B:AIRE.0000045502.10941.a9}.
\bibitem[{Hus{\'{a}}k and Ka{\v{s}}par(2019)}]{Husak2019}
\bibinfo{author}{M.~Hus{\'{a}}k}, \bibinfo{author}{J.~Ka{\v{s}}par},
\newblock \bibinfo{title}{{AIDA framework: Real-time correlation and prediction
  of intrusion detection alerts}},
\newblock in: \bibinfo{booktitle}{ACM International Conference Proceeding
  Series}, \bibinfo{publisher}{Association for Computing Machinery},
  \bibinfo{address}{New York, New York, USA}, \bibinfo{year}{2019}, pp.
  \bibinfo{pages}{1--8}. \URLprefix
  \url{http://dl.acm.org/citation.cfm?doid=3339252.3340513}.
  \DOIprefix\doi{10.1145/3339252.3340513}.
\bibitem[{Ficco(2013)}]{Ficco2013}
\bibinfo{author}{M.~Ficco},
\newblock \bibinfo{title}{{Security event correlation approach for cloud
  computing}},
\newblock \bibinfo{journal}{International Journal of High Performance Computing
  and Networking} \bibinfo{volume}{7} (\bibinfo{year}{2013})
  \bibinfo{pages}{173}. \URLprefix
  \url{http://www.inderscience.com/link.php?id=56525}.
  \DOIprefix\doi{10.1504/IJHPCN.2013.056525}.
\bibitem[{Buneman et~al.(2001)Buneman, Khanna, and Wang-Chiew}]{buneman2001and}
\bibinfo{author}{P.~Buneman}, \bibinfo{author}{S.~Khanna},
  \bibinfo{author}{T.~Wang-Chiew},
\newblock \bibinfo{title}{Why and where: A characterization of data
  provenance},
\newblock in: \bibinfo{booktitle}{International conference on database theory},
  \bibinfo{organization}{Springer}, \bibinfo{year}{2001}, pp.
  \bibinfo{pages}{316--330}.
\bibitem[{Woodruff and Stonebraker(1997)}]{woodruff1997supporting}
\bibinfo{author}{A.~Woodruff}, \bibinfo{author}{M.~Stonebraker},
\newblock \bibinfo{title}{Supporting fine-grained data lineage in a database
  visualization environment},
\newblock in: \bibinfo{booktitle}{Proceedings 13th International Conference on
  Data Engineering}, \bibinfo{organization}{IEEE}, \bibinfo{year}{1997}, pp.
  \bibinfo{pages}{91--102}.
\bibitem[{Greenwood et~al.(2003)Greenwood, Goble, Stevens, Zhao, Addis, Marvin,
  Moreau, and Oinn}]{greenwood2003provenance}
\bibinfo{author}{M.~Greenwood}, \bibinfo{author}{C.~Goble},
  \bibinfo{author}{R.~Stevens}, \bibinfo{author}{J.~Zhao},
  \bibinfo{author}{M.~Addis}, \bibinfo{author}{D.~Marvin},
  \bibinfo{author}{L.~Moreau}, \bibinfo{author}{T.~Oinn},
\newblock \bibinfo{title}{Provenance of e-science experiments-experience from
  bioinformatics}  (\bibinfo{year}{2003}).
\bibitem[{Miles et~al.(2007)Miles, Groth, Branco, and
  Moreau}]{miles2007requirements}
\bibinfo{author}{S.~Miles}, \bibinfo{author}{P.~Groth},
  \bibinfo{author}{M.~Branco}, \bibinfo{author}{L.~Moreau},
\newblock \bibinfo{title}{The requirements of using provenance in e-science
  experiments},
\newblock \bibinfo{journal}{Journal of Grid Computing} \bibinfo{volume}{5}
  (\bibinfo{year}{2007}) \bibinfo{pages}{1--25}.
\bibitem[{Naughton et~al.(2009)Naughton, Bland, Vallee, Engelmann, and
  Scott}]{naughton2009fault}
\bibinfo{author}{T.~Naughton}, \bibinfo{author}{W.~Bland},
  \bibinfo{author}{G.~Vallee}, \bibinfo{author}{C.~Engelmann},
  \bibinfo{author}{S.~L. Scott},
\newblock \bibinfo{title}{Fault injection framework for system resilience
  evaluation: fake faults for finding future failures},
\newblock in: \bibinfo{booktitle}{Proceedings of the 2009 workshop on
  Resiliency in high performance}, \bibinfo{year}{2009}, pp.
  \bibinfo{pages}{23--28}.
\bibitem[{Simmhan et~al.(2005)Simmhan, Plale, and Gannon}]{simmhan2005survey}
\bibinfo{author}{Y.~L. Simmhan}, \bibinfo{author}{B.~Plale},
  \bibinfo{author}{D.~Gannon},
\newblock \bibinfo{title}{A survey of data provenance in e-science},
\newblock \bibinfo{journal}{ACM Sigmod Record} \bibinfo{volume}{34}
  (\bibinfo{year}{2005}) \bibinfo{pages}{31--36}.
\bibitem[{Freire et~al.(2008)Freire, Koop, Santos, and
  Silva}]{freire_provenance_2008}
\bibinfo{author}{J.~Freire}, \bibinfo{author}{D.~Koop},
  \bibinfo{author}{E.~Santos}, \bibinfo{author}{C.~Silva},
\newblock \bibinfo{title}{Provenance for {Computational} {Tasks}: {A}
  {Survey}},
\newblock \bibinfo{journal}{Computing in Science \& Engineering}
  \bibinfo{volume}{10} (\bibinfo{year}{2008}) \bibinfo{pages}{11--21}.
  \URLprefix \url{http://ieeexplore.ieee.org/document/4488060/}.
  \DOIprefix\doi{10.1109/MCSE.2008.79}.
\bibitem[{Zafar et~al.(2017)Zafar, Khan, Suhail, Ahmed, Hameed, Khan, Jabeen,
  and Anjum}]{zafar_trustworthy_2017}
\bibinfo{author}{F.~Zafar}, \bibinfo{author}{A.~Khan},
  \bibinfo{author}{S.~Suhail}, \bibinfo{author}{I.~Ahmed},
  \bibinfo{author}{K.~Hameed}, \bibinfo{author}{H.~M. Khan},
  \bibinfo{author}{F.~Jabeen}, \bibinfo{author}{A.~Anjum},
\newblock \bibinfo{title}{Trustworthy data: {A} survey, taxonomy and future
  trends of secure provenance schemes},
\newblock \bibinfo{journal}{Journal of Network and Computer Applications}
  \bibinfo{volume}{94} (\bibinfo{year}{2017}) \bibinfo{pages}{50--68}.
  \DOIprefix\doi{10.1016/j.jnca.2017.06.003}.
\bibitem[{Herschel(2017)}]{herschel_survey_nodate}
\bibinfo{author}{M.~Herschel},
\newblock \bibinfo{title}{A survey on provenance: {What} for? {What} form?
  {What} from?}  (\bibinfo{year}{2017}) \bibinfo{pages}{26}.
\bibitem[{Venkatasubramanian et~al.(2003)Venkatasubramanian, Hayes, and
  Murray}]{Venkatasubramanian2003}
\bibinfo{author}{R.~Venkatasubramanian}, \bibinfo{author}{J.~P. Hayes},
  \bibinfo{author}{B.~T. Murray},
\newblock \bibinfo{title}{{Low-cost on-line fault detection using control flow
  assertions}},
\newblock in: \bibinfo{booktitle}{Proceedings - 9th IEEE International On-Line
  Testing Symposium, IOLTS 2003}, \bibinfo{publisher}{Institute of Electrical
  and Electronics Engineers Inc.}, \bibinfo{year}{2003}, pp.
  \bibinfo{pages}{137--143}. \DOIprefix\doi{10.1109/OLT.2003.1214380}.
\bibitem[{Petroni and Hicks(2007)}]{Petroni2007}
\bibinfo{author}{N.~L. Petroni}, \bibinfo{author}{M.~Hicks},
\newblock \bibinfo{title}{{Automated detection of persistent kernel
  control-flow attacks}},
\newblock in: \bibinfo{booktitle}{Proceedings of the ACM Conference on Computer
  and Communications Security}, \bibinfo{publisher}{ACM Press},
  \bibinfo{address}{New York, New York, USA}, \bibinfo{year}{2007}, pp.
  \bibinfo{pages}{103--115}. \URLprefix
  \url{http://portal.acm.org/citation.cfm?doid=1315245.1315260}.
  \DOIprefix\doi{10.1145/1315245.1315260}.
\bibitem[{Ndichu et~al.(2019)Ndichu, Kim, Ozawa, Misu, and
  Makishima}]{Ndichu2019}
\bibinfo{author}{S.~Ndichu}, \bibinfo{author}{S.~Kim},
  \bibinfo{author}{S.~Ozawa}, \bibinfo{author}{T.~Misu},
  \bibinfo{author}{K.~Makishima},
\newblock \bibinfo{title}{{A machine learning approach to detection of
  JavaScript-based attacks using AST features and paragraph vectors}},
\newblock \bibinfo{journal}{Applied Soft Computing Journal}
  \bibinfo{volume}{84} (\bibinfo{year}{2019}) \bibinfo{pages}{105721}.
  \DOIprefix\doi{10.1016/j.asoc.2019.105721}.
\bibitem[{Li et~al.(2019)Li, Chen, Chen, Zhu, Xiong, and Yang}]{Li2019}
\bibinfo{author}{Z.~Li}, \bibinfo{author}{Y.~Chen}, \bibinfo{author}{Q.~Chen},
  \bibinfo{author}{T.~Zhu}, \bibinfo{author}{C.~Xiong},
  \bibinfo{author}{H.~Yang},
\newblock \bibinfo{title}{{Effective and light-weight deobfuscation and
  semantic-aware attack detection for powershell scripts}},
\newblock in: \bibinfo{booktitle}{Proceedings of the ACM Conference on Computer
  and Communications Security}, \bibinfo{year}{2019}.
  \DOIprefix\doi{10.1145/3319535.3363187}.
\bibitem[{Mu{\~{n}}oz-Gonz{\'{a}}lez et~al.(2016)Mu{\~{n}}oz-Gonz{\'{a}}lez,
  Sgandurra, Paudice, and Lupu}]{Munoz-Gonzalez2016}
\bibinfo{author}{L.~Mu{\~{n}}oz-Gonz{\'{a}}lez},
  \bibinfo{author}{D.~Sgandurra}, \bibinfo{author}{A.~Paudice},
  \bibinfo{author}{E.~C. Lupu},
\newblock \bibinfo{title}{{Efficient Attack Graph Analysis through Approximate
  Inference}},
\newblock \bibinfo{journal}{ACM Transactions on Privacy and Security}
  \bibinfo{volume}{20} (\bibinfo{year}{2016}). \URLprefix
  \url{http://arxiv.org/abs/1606.07025}.
  \href{http://arxiv.org/abs/1606.07025}{{\tt arXiv:1606.07025}}.
\bibitem[{Wu et~al.(2012)Wu, Yin, and Guo}]{Wu2012}
\bibinfo{author}{J.~Wu}, \bibinfo{author}{L.~Yin}, \bibinfo{author}{Y.~Guo},
\newblock \bibinfo{title}{{Cyber attacks prediction model based on Bayesian
  network}},
\newblock in: \bibinfo{booktitle}{Proceedings of the International Conference
  on Parallel and Distributed Systems - ICPADS}, \bibinfo{year}{2012}, pp.
  \bibinfo{pages}{730--731}. \DOIprefix\doi{10.1109/ICPADS.2012.117}.
\bibitem[{Frigault and Wang(2008)}]{Frigault2008}
\bibinfo{author}{M.~Frigault}, \bibinfo{author}{L.~Wang},
\newblock \bibinfo{title}{{Measuring network security using bayesian
  network-based attack graphs}},
\newblock in: \bibinfo{booktitle}{Proceedings - International Computer Software
  and Applications Conference}, \bibinfo{year}{2008}, pp.
  \bibinfo{pages}{698--703}. \DOIprefix\doi{10.1109/COMPSAC.2008.88}.
\bibitem[{Xu and Nygard(2006)}]{Xu2006}
\bibinfo{author}{D.~Xu}, \bibinfo{author}{K.~Nygard},
\newblock \bibinfo{title}{{Threat-driven modeling and verification of secure
  software using aspect-oriented Petri nets}},
\newblock \bibinfo{journal}{IEEE Transactions on Software Engineering}
  \bibinfo{volume}{32} (\bibinfo{year}{2006}) \bibinfo{pages}{265--278}.
  \DOIprefix\doi{10.1109/tse.2006.40}.
\bibitem[{con(0203)}]{connectedpapers}
\bibinfo{title}{Connected papers}, \bibinfo{year}{2020.3}. \URLprefix
  \url{https://www.connectedpapers.com/}.
\bibitem[{Pasquier et~al.(2017)Pasquier, Han, Goldstein, Moyer, Eyers, Seltzer,
  and Bacon}]{pasquier_practical_2017}
\bibinfo{author}{T.~Pasquier}, \bibinfo{author}{X.~Han},
  \bibinfo{author}{M.~Goldstein}, \bibinfo{author}{T.~Moyer},
  \bibinfo{author}{D.~Eyers}, \bibinfo{author}{M.~Seltzer},
  \bibinfo{author}{J.~Bacon},
\newblock \bibinfo{title}{Practical whole-system provenance capture},
\newblock in: \bibinfo{booktitle}{Proceedings of the 2017 {Symposium} on
  {Cloud} {Computing}}, {SoCC} '17, \bibinfo{publisher}{Association for
  Computing Machinery}, \bibinfo{address}{Santa Clara, California},
  \bibinfo{year}{2017}, pp. \bibinfo{pages}{405--418}. \URLprefix
  \url{https://doi.org/10.1145/3127479.3129249}.
  \DOIprefix\doi{10.1145/3127479.3129249}.
\bibitem[{Ma et~al.(2017)Ma, Zhai, Wang, Lee, Zhang, and Xu}]{ma_mpi:_2017}
\bibinfo{author}{S.~Ma}, \bibinfo{author}{J.~Zhai}, \bibinfo{author}{F.~Wang},
  \bibinfo{author}{K.~H. Lee}, \bibinfo{author}{X.~Zhang},
  \bibinfo{author}{D.~Xu},
\newblock \bibinfo{title}{{MPI}: {Multiple} {Perspective} {Attack}
  {Investigation} with {Semantic} {Aware} {Execution} {Partitioning}},
\newblock in: \bibinfo{booktitle}{26th {USENIX} {Security} {Symposium}
  ({USENIX} {Security} 17)}, \bibinfo{publisher}{USENIX Association},
  \bibinfo{address}{Vancouver, BC}, \bibinfo{year}{2017}, pp.
  \bibinfo{pages}{1111--1128}.
\bibitem[{Lee et~al.(2013)Lee, Zhang, and Xu}]{lee_high_nodate}
\bibinfo{author}{K.~H. Lee}, \bibinfo{author}{X.~Zhang},
  \bibinfo{author}{D.~Xu},
\newblock \bibinfo{title}{High {Accuracy} {Attack} {Provenance} via
  {Binary}-based {Execution} {Partition}},
\newblock in: \bibinfo{booktitle}{NDSS}, \bibinfo{year}{2013},
  p.~\bibinfo{pages}{16}.
\bibitem[{Yang et~al.(2020)Yang, Ma, Xu, Zhang, and Chen}]{yanguiscope}
\bibinfo{author}{R.~Yang}, \bibinfo{author}{S.~Ma}, \bibinfo{author}{H.~Xu},
  \bibinfo{author}{X.~Zhang}, \bibinfo{author}{Y.~Chen},
\newblock \bibinfo{title}{Uiscope: Accurate, instrumentation-free, and visible
  attack investigation for gui applications}  (\bibinfo{year}{2020}).
\bibitem[{Ma et~al.(2016)Ma, Zhang, and Xu}]{ma_protracer:_2016}
\bibinfo{author}{S.~Ma}, \bibinfo{author}{X.~Zhang}, \bibinfo{author}{D.~Xu},
\newblock \bibinfo{title}{{ProTracer}: {Towards} {Practical} {Provenance}
  {Tracing} by {Alternating} {Between} {Logging} and {Tainting}},
\newblock \bibinfo{publisher}{Internet Society}, \bibinfo{year}{2016}.
  \DOIprefix\doi{10.14722/ndss.2016.23350}.
\bibitem[{Kwon et~al.(2016)Kwon, Kim, Sumner, Kim, Saltaformaggio, Zhang, and
  Xu}]{kwon_ldx:_2016}
\bibinfo{author}{Y.~Kwon}, \bibinfo{author}{D.~Kim}, \bibinfo{author}{W.~N.
  Sumner}, \bibinfo{author}{K.~Kim}, \bibinfo{author}{B.~Saltaformaggio},
  \bibinfo{author}{X.~Zhang}, \bibinfo{author}{D.~Xu},
\newblock \bibinfo{title}{{LDX}: {Causality} {Inference} by {Lightweight}
  {Dual} {Execution}},
\newblock in: \bibinfo{booktitle}{Proceedings of the {Twenty}-{First}
  {International} {Conference} on {Architectural} {Support} for {Programming}
  {Languages} and {Operating} {Systems}}, {ASPLOS} '16,
  \bibinfo{publisher}{ACM}, \bibinfo{address}{New York, NY, USA},
  \bibinfo{year}{2016}, pp. \bibinfo{pages}{503--515}. \URLprefix
  \url{http://doi.acm.org/10.1145/2872362.2872395}.
  \DOIprefix\doi{10.1145/2872362.2872395}.
\bibitem[{Kwon et~al.(2018)Kwon, Wang, Wang, Lee, Lee, Ma, Zhang, Xu, Jha,
  Ciocarlie, Gehani, and Yegneswaran}]{kwon_mci_2018}
\bibinfo{author}{Y.~Kwon}, \bibinfo{author}{F.~Wang},
  \bibinfo{author}{W.~Wang}, \bibinfo{author}{K.~H. Lee},
  \bibinfo{author}{W.-C. Lee}, \bibinfo{author}{S.~Ma},
  \bibinfo{author}{X.~Zhang}, \bibinfo{author}{D.~Xu},
  \bibinfo{author}{S.~Jha}, \bibinfo{author}{G.~Ciocarlie},
  \bibinfo{author}{A.~Gehani}, \bibinfo{author}{V.~Yegneswaran},
\newblock \bibinfo{title}{{MCI} : {Modeling}-based {Causality} {Inference} in
  {Audit} {Logging} for {Attack} {Investigation}},
\newblock \bibinfo{publisher}{Internet Society}, \bibinfo{year}{2018}.
  \DOIprefix\doi{10.14722/ndss.2018.23306}.
\bibitem[{Hassan et~al.(2018)Hassan, Lemay, Aguse, Bates, and
  Moyer}]{hassan_towards_2018}
\bibinfo{author}{W.~U. Hassan}, \bibinfo{author}{M.~Lemay},
  \bibinfo{author}{N.~Aguse}, \bibinfo{author}{A.~Bates},
  \bibinfo{author}{T.~Moyer},
\newblock \bibinfo{title}{Towards {Scalable} {Cluster} {Auditing} through
  {Grammatical} {Inference} over {Provenance} {Graphs}},
\newblock in: \bibinfo{booktitle}{Proceedings 2018 {Network} and {Distributed}
  {System} {Security} {Symposium}}, \bibinfo{publisher}{Internet Society},
  \bibinfo{address}{San Diego, CA}, \bibinfo{year}{2018}.
  \DOIprefix\doi{10.14722/ndss.2018.23141}.
\bibitem[{Ji et~al.(2017)Ji, Lee, Downing, Wang, Fazzini, Kim, Orso, and
  Lee}]{ji_rain:_2017}
\bibinfo{author}{Y.~Ji}, \bibinfo{author}{S.~Lee},
  \bibinfo{author}{E.~Downing}, \bibinfo{author}{W.~Wang},
  \bibinfo{author}{M.~Fazzini}, \bibinfo{author}{T.~Kim},
  \bibinfo{author}{A.~Orso}, \bibinfo{author}{W.~Lee},
\newblock \bibinfo{title}{{RAIN}: {Refinable} {Attack} {Investigation} with
  {On}-demand {Inter}-{Process} {Information} {Flow} {Tracking}},
\newblock in: \bibinfo{booktitle}{Proceedings of the 2017 {ACM} {SIGSAC}
  {Conference} on {Computer} and {Communications} {Security} - {CCS} '17},
  \bibinfo{publisher}{ACM Press}, \bibinfo{address}{Dallas, Texas, USA},
  \bibinfo{year}{2017}, pp. \bibinfo{pages}{377--390}. \URLprefix
  \url{http://dl.acm.org/citation.cfm?doid=3133956.3134045}.
  \DOIprefix\doi{10.1145/3133956.3134045}.
\bibitem[{Kemerlis et~al.(2012)Kemerlis, Portokalidis, Jee, and
  Keromytis}]{kemerlis_libdft:_2012}
\bibinfo{author}{V.~P. Kemerlis}, \bibinfo{author}{G.~Portokalidis},
  \bibinfo{author}{K.~Jee}, \bibinfo{author}{A.~D. Keromytis},
\newblock \bibinfo{title}{Libdft: {Practical} {Dynamic} {Data} {Flow}
  {Tracking} for {Commodity} {Systems}},
\newblock in: \bibinfo{booktitle}{Proceedings of the 8th {ACM}
  {SIGPLAN}/{SIGOPS} {Conference} on {Virtual} {Execution} {Environments}},
  {VEE} '12, \bibinfo{publisher}{ACM}, \bibinfo{address}{New York, NY, USA},
  \bibinfo{year}{2012}, pp. \bibinfo{pages}{121--132}. \URLprefix
  \url{http://doi.acm.org/10.1145/2151024.2151042}.
  \DOIprefix\doi{10.1145/2151024.2151042}.
\bibitem[{Ji et~al.(2018)Ji, Lee, Fazzini, Allen, Downing, Kim, Orso, and
  Lee}]{yang_enabling_2018}
\bibinfo{author}{Y.~Ji}, \bibinfo{author}{S.~Lee},
  \bibinfo{author}{M.~Fazzini}, \bibinfo{author}{J.~Allen},
  \bibinfo{author}{E.~Downing}, \bibinfo{author}{T.~Kim},
  \bibinfo{author}{A.~Orso}, \bibinfo{author}{W.~Lee},
\newblock \bibinfo{title}{Enabling refinable cross-host attack investigation
  with efficient data flow tagging and tracking},
\newblock in: \bibinfo{booktitle}{27th {USENIX} Security Symposium ({USENIX}
  Security 18)}, \bibinfo{publisher}{{USENIX} Association},
  \bibinfo{address}{Baltimore, MD}, \bibinfo{year}{2018}, pp.
  \bibinfo{pages}{1705--1722}.
\bibitem[{Hassan et~al.(2020)Hassan, Noureddine, Datta, and
  Bates}]{hassan_omegalog_nodate}
\bibinfo{author}{W.~U. Hassan}, \bibinfo{author}{M.~A. Noureddine},
  \bibinfo{author}{P.~Datta}, \bibinfo{author}{A.~Bates},
\newblock \bibinfo{title}{{OmegaLog}: {High}-{Fidelity} {Attack}
  {Investigation} via {Transparent} {Multi}-layer {Log} {Analysis}},
\newblock \bibinfo{year}{2020}, p.~\bibinfo{pages}{16}.
\bibitem[{fus(2020)}]{fuse}
\bibinfo{title}{Linux fuse}, \bibinfo{year}{2020}. \URLprefix
  \url{https://github.com/libfuse/libfuse}.
\bibitem[{Schaufler(2016)}]{lsm}
\bibinfo{author}{C.~Schaufler}, \bibinfo{title}{Lsm: Stacking for major
  security modules}, \bibinfo{year}{2016}. \URLprefix
  \url{https://lwn.net/Articles/697259/}.
\bibitem[{net(2020)}]{netfilter}
\bibinfo{title}{Netfilter}, \bibinfo{year}{2020}. \URLprefix
  \url{https://www.netfilter.org/}.
\bibitem[{OSX(2020)}]{OSXFuse}
\bibinfo{title}{Osxfuse}, \bibinfo{year}{2020}. \URLprefix
  \url{https://osxfuse.github.io/}.
\bibitem[{pro(2020)}]{prov-dm}
\bibinfo{title}{W3c prov-dm}, \bibinfo{year}{2020}. \URLprefix
  \url{https://dvcs.w3.org/hg/prov/raw\-file/default/model/working\-copy/prov\-dm\-issue\-450.html}.
\bibitem[{Clause et~al.(2007)Clause, Li, and Orso}]{clause2007dytan}
\bibinfo{author}{J.~Clause}, \bibinfo{author}{W.~Li},
  \bibinfo{author}{A.~Orso},
\newblock \bibinfo{title}{Dytan: a generic dynamic taint analysis framework},
\newblock in: \bibinfo{booktitle}{Proceedings of the 2007 international
  symposium on Software testing and analysis}, \bibinfo{year}{2007}, pp.
  \bibinfo{pages}{196--206}.
\bibitem[{Enck et~al.(2014)Enck, Gilbert, Han, Tendulkar, Chun, Cox, Jung,
  McDaniel, and Sheth}]{enck2014taintdroid}
\bibinfo{author}{W.~Enck}, \bibinfo{author}{P.~Gilbert},
  \bibinfo{author}{S.~Han}, \bibinfo{author}{V.~Tendulkar},
  \bibinfo{author}{B.-G. Chun}, \bibinfo{author}{L.~P. Cox},
  \bibinfo{author}{J.~Jung}, \bibinfo{author}{P.~McDaniel},
  \bibinfo{author}{A.~N. Sheth},
\newblock \bibinfo{title}{Taintdroid: an information-flow tracking system for
  realtime privacy monitoring on smartphones},
\newblock \bibinfo{journal}{ACM Transactions on Computer Systems (TOCS)}
  \bibinfo{volume}{32} (\bibinfo{year}{2014}) \bibinfo{pages}{1--29}.
\bibitem[{Xu et~al.(2006)Xu, Bhatkar, and Sekar}]{xu2006taint}
\bibinfo{author}{W.~Xu}, \bibinfo{author}{S.~Bhatkar},
  \bibinfo{author}{R.~Sekar},
\newblock \bibinfo{title}{Taint-enhanced policy enforcement: A practical
  approach to defeat a wide range of attacks.},
\newblock in: \bibinfo{booktitle}{USENIX Security Symposium},
  \bibinfo{year}{2006}, pp. \bibinfo{pages}{121--136}.
\bibitem[{gra(2020)}]{graph_database}
\bibinfo{title}{Graph database}, \bibinfo{year}{2020}. \URLprefix
  \url{en.wikipedia.org/wiki/Graph\_database}.
\bibitem[{Han et~al.(2020)Han, Pasquier, Bates, Mickens, and
  Seltzer}]{han_unicorn_2020}
\bibinfo{author}{X.~Han}, \bibinfo{author}{T.~Pasquier},
  \bibinfo{author}{A.~Bates}, \bibinfo{author}{J.~Mickens},
  \bibinfo{author}{M.~Seltzer},
\newblock \bibinfo{title}{{UNICORN}: {Runtime} {Provenance}-{Based} {Detector}
  for {Advanced} {Persistent} {Threats}},
\newblock in: \bibinfo{booktitle}{{arXiv} preprint {arXiv}:2001.01525},
  \bibinfo{year}{2020}.
\bibitem[{Hossain et~al.(2018)Hossain, Wang, Sekar, and
  Stoller}]{hossain_dependence-preserving_2018}
\bibinfo{author}{M.~N. Hossain}, \bibinfo{author}{J.~Wang},
  \bibinfo{author}{R.~Sekar}, \bibinfo{author}{S.~D. Stoller},
\newblock \bibinfo{title}{Dependence-preserving data compaction for scalable
  forensic analysis},
\newblock \bibinfo{year}{2018}, pp. \bibinfo{pages}{1723--1740}.
\bibitem[{noa(2016)}]{noauthor_provenance-aware_2016}
\bibinfo{title}{Provenance-aware versioned dataworkspaces},
\newblock in: \bibinfo{booktitle}{8th {USENIX} Workshop on the Theory and
  Practice of Provenance ({TaPP} 16)}, \bibinfo{publisher}{{USENIX}
  Association}, \bibinfo{year}{2016}. \URLprefix
  \url{https://www.usenix.org/conference/tapp16/workshop\-program/pressentation/niu}.
\bibitem[{Chavan et~al.(2015)Chavan, Huang, Deshpande, Elmore, Madden, and
  Parameswaran}]{chavan_towards_2015}
\bibinfo{author}{A.~Chavan}, \bibinfo{author}{S.~Huang},
  \bibinfo{author}{A.~Deshpande}, \bibinfo{author}{A.~Elmore},
  \bibinfo{author}{S.~Madden}, \bibinfo{author}{A.~Parameswaran},
\newblock \bibinfo{title}{Towards a unified query language for provenance and
  versioning},
\newblock in: \bibinfo{booktitle}{7th {USENIX} Workshop on the Theory and
  Practice of Provenance ({TaPP} 15)}, \bibinfo{publisher}{{USENIX}
  Association}, \bibinfo{year}{2015}. \URLprefix
  \url{https://www.usenix.org/conference/tapp15/workshop\-program/presentation/chavan}.
\bibitem[{Chapman et~al.(2008)Chapman, Jagadish, and
  Ramanan}]{chapman2008efficient}
\bibinfo{author}{A.~P. Chapman}, \bibinfo{author}{H.~V. Jagadish},
  \bibinfo{author}{P.~Ramanan},
\newblock \bibinfo{title}{Efficient provenance storage},
\newblock in: \bibinfo{booktitle}{Proceedings of the 2008 ACM SIGMOD
  international conference on Management of data}, \bibinfo{year}{2008}, pp.
  \bibinfo{pages}{993--1006}.
\bibitem[{Xie et~al.(2012)Xie, Feng, Tan, Chen, Muniswamy-Reddy, Li, and
  Long}]{xie2012hybrid}
\bibinfo{author}{Y.~Xie}, \bibinfo{author}{D.~Feng}, \bibinfo{author}{Z.~Tan},
  \bibinfo{author}{L.~Chen}, \bibinfo{author}{K.-K. Muniswamy-Reddy},
  \bibinfo{author}{Y.~Li}, \bibinfo{author}{D.~D. Long},
\newblock \bibinfo{title}{A hybrid approach for efficient provenance storage},
\newblock in: \bibinfo{booktitle}{Proceedings of the 21st ACM international
  conference on Information and knowledge management}, \bibinfo{year}{2012},
  pp. \bibinfo{pages}{1752--1756}.
\bibitem[{Lee et~al.(2013)Lee, Zhang, and Xu}]{lee_loggc:_2013}
\bibinfo{author}{K.~H. Lee}, \bibinfo{author}{X.~Zhang},
  \bibinfo{author}{D.~Xu},
\newblock \bibinfo{title}{{LogGC}: {Garbage} {Collecting} {Audit} {Log}},
\newblock in: \bibinfo{booktitle}{Proceedings of the 2013 {ACM} {SIGSAC}
  {Conference} on {Computer} \& {Communications} {Security}}, {CCS} '13,
  \bibinfo{publisher}{ACM}, \bibinfo{address}{New York, NY, USA},
  \bibinfo{year}{2013}, pp. \bibinfo{pages}{1005--1016}. \URLprefix
  \url{http://doi.acm.org/10.1145/2508859.2516731}.
  \DOIprefix\doi{10.1145/2508859.2516731}.
\bibitem[{Tang et~al.(2018)Tang, Li, Li, Li, Zhang, Jee, Xiao, Wu, Rhee, and
  Xu}]{tang_nodemerge:_2018}
\bibinfo{author}{Y.~Tang}, \bibinfo{author}{Q.~Li}, \bibinfo{author}{D.~Li},
  \bibinfo{author}{Z.~Li}, \bibinfo{author}{M.~Zhang},
  \bibinfo{author}{K.~Jee}, \bibinfo{author}{X.~Xiao}, \bibinfo{author}{Z.~Wu},
  \bibinfo{author}{J.~Rhee}, \bibinfo{author}{F.~Xu},
\newblock \bibinfo{title}{{NodeMerge}: {Template} {Based} {Efficient} {Data}
  {Reduction} {For} {Big}-{Data} {Causality} {Analysis}},
\newblock in: \bibinfo{booktitle}{Proceedings of the 2018 {ACM} {SIGSAC}
  {Conference} on {Computer} and {Communications} {Security} - {CCS} '18},
  \bibinfo{publisher}{ACM Press}, \bibinfo{address}{Toronto, Canada},
  \bibinfo{year}{2018}, pp. \bibinfo{pages}{1324--1337}. \URLprefix
  \url{http://dl.acm.org/citation.cfm?doid=3243734.3243763}.
  \DOIprefix\doi{10.1145/3243734.3243763}.
\bibitem[{Gao et~al.(2018{\natexlab{a}})Gao, Xiao, Li, Jee, Xu, Kulkarni, and
  Mittal}]{gao_aiql:_2018}
\bibinfo{author}{P.~Gao}, \bibinfo{author}{X.~Xiao}, \bibinfo{author}{Z.~Li},
  \bibinfo{author}{K.~Jee}, \bibinfo{author}{F.~Xu}, \bibinfo{author}{S.~R.
  Kulkarni}, \bibinfo{author}{P.~Mittal},
\newblock \bibinfo{title}{{AIQL}: {Enabling} {Efficient} {Attack}
  {Investigation} from {System} {Monitoring} {Data}},
\newblock in: \bibinfo{booktitle}{{arXiv}:1806.02290 [cs]},
  \bibinfo{year}{2018}{\natexlab{a}}. \URLprefix
  \url{http://arxiv.org/abs/1806.02290}, \bibinfo{note}{arXiv: 1806.02290}.
\bibitem[{Gao et~al.(2018{\natexlab{b}})Gao, Xiao, Li, Li, Jee, Wu, Kim,
  Kulkarni, and Mittal}]{gao_saql:_2018}
\bibinfo{author}{P.~Gao}, \bibinfo{author}{X.~Xiao}, \bibinfo{author}{D.~Li},
  \bibinfo{author}{Z.~Li}, \bibinfo{author}{K.~Jee}, \bibinfo{author}{Z.~Wu},
  \bibinfo{author}{C.~H. Kim}, \bibinfo{author}{S.~R. Kulkarni},
  \bibinfo{author}{P.~Mittal},
\newblock \bibinfo{title}{{SAQL}: {A} {Stream}-based {Query} {System} for
  {Real}-{Time} {Abnormal} {System} {Behavior} {Detection}},
\newblock in: \bibinfo{booktitle}{27th {USENIX} {Security} {Symposium}
  ({USENIX} {Security} 18)}, \bibinfo{publisher}{USENIX Association},
  \bibinfo{address}{Baltimore, MD}, \bibinfo{year}{2018}{\natexlab{b}}, pp.
  \bibinfo{pages}{639--656}.
\bibitem[{Shu et~al.(2018)Shu, Araujo, Schales, Stoecklin, Jang, Huang, and
  Rao}]{shu_threat_2018}
\bibinfo{author}{X.~Shu}, \bibinfo{author}{F.~Araujo}, \bibinfo{author}{D.~L.
  Schales}, \bibinfo{author}{M.~P. Stoecklin}, \bibinfo{author}{J.~Jang},
  \bibinfo{author}{H.~Huang}, \bibinfo{author}{J.~R. Rao},
\newblock \bibinfo{title}{Threat {Intelligence} {Computing}},
\newblock in: \bibinfo{booktitle}{Proceedings of the 2018 {ACM} {SIGSAC}
  {Conference} on {Computer} and {Communications} {Security}}, {CCS} '18,
  \bibinfo{publisher}{ACM}, \bibinfo{address}{New York, NY, USA},
  \bibinfo{year}{2018}, pp. \bibinfo{pages}{1883--1898}. \URLprefix
  \url{http://doi.acm.org/10.1145/3243734.3243829}.
  \DOIprefix\doi{10.1145/3243734.3243829}.
\bibitem[{Pasquier et~al.(2018)Pasquier, Han, Moyer, Bates, Hermant, Eyers,
  Bacon, and Seltzer}]{pasquier_runtime_2018}
\bibinfo{author}{T.~Pasquier}, \bibinfo{author}{X.~Han},
  \bibinfo{author}{T.~Moyer}, \bibinfo{author}{A.~Bates},
  \bibinfo{author}{O.~Hermant}, \bibinfo{author}{D.~Eyers},
  \bibinfo{author}{J.~Bacon}, \bibinfo{author}{M.~Seltzer},
\newblock \bibinfo{title}{Runtime {Analysis} of {Whole}-{System} {Provenance}},
\newblock in: \bibinfo{booktitle}{Proceedings of the 2018 {ACM} {SIGSAC}
  {Conference} on {Computer} and {Communications} {Security} - {CCS} '18},
  \bibinfo{publisher}{ACM Press}, \bibinfo{address}{Toronto, Canada},
  \bibinfo{year}{2018}, pp. \bibinfo{pages}{1601--1616}. \URLprefix
  \url{http://dl.acm.org/citation.cfm?doid=3243734.3243776}.
  \DOIprefix\doi{10.1145/3243734.3243776}.
\bibitem[{King and Chen(2003)}]{king_backtracking_2003}
\bibinfo{author}{S.~T. King}, \bibinfo{author}{P.~M. Chen},
\newblock \bibinfo{title}{Backtracking {Intrusions}},
\newblock in: \bibinfo{booktitle}{Proceedings of the {Nineteenth} {ACM}
  {Symposium} on {Operating} {Systems} {Principles}}, {SOSP} '03,
  \bibinfo{publisher}{ACM}, \bibinfo{address}{New York, NY, USA},
  \bibinfo{year}{2003}, pp. \bibinfo{pages}{223--236}. \URLprefix
  \url{http://doi.acm.org/10.1145/945445.945467}.
  \DOIprefix\doi{10.1145/945445.945467}.
\bibitem[{Sid(2020)}]{Siddhi}
\bibinfo{title}{Siddhi complex event processing engine}, \bibinfo{year}{2020}.
  \URLprefix \url{https://github.com/wso2/siddhi}.
\bibitem[{Mit(2020)}]{Mitre}
\bibinfo{title}{Mitre att\&ck metrics}, \bibinfo{year}{2020}. \URLprefix
  \url{https://attack.mitre.org/}.
\bibitem[{Pei et~al.(2016)Pei, Gu, Saltaformaggio, Ma, Wang, Zhang, Si, Zhang,
  and Xu}]{pei_hercule:_2016}
\bibinfo{author}{K.~Pei}, \bibinfo{author}{Z.~Gu},
  \bibinfo{author}{B.~Saltaformaggio}, \bibinfo{author}{S.~Ma},
  \bibinfo{author}{F.~Wang}, \bibinfo{author}{Z.~Zhang},
  \bibinfo{author}{L.~Si}, \bibinfo{author}{X.~Zhang}, \bibinfo{author}{D.~Xu},
\newblock \bibinfo{title}{{HERCULE}: {Attack} {Story} {Reconstruction} via
  {Community} {Discovery} on {Correlated} {Log} {Graph}},
\newblock in: \bibinfo{booktitle}{Proceedings of the {32Nd} {Annual}
  {Conference} on {Computer} {Security} {Applications}}, {ACSAC} '16,
  \bibinfo{publisher}{ACM}, \bibinfo{address}{New York, NY, USA},
  \bibinfo{year}{2016}, pp. \bibinfo{pages}{583--595}. \URLprefix
  \url{http://doi.acm.org/10.1145/2991079.2991122}.
  \DOIprefix\doi{10.1145/2991079.2991122}.
\bibitem[{Liu et~al.(2019)Liu, Wen, Zhang, Jiang, Xing, and
  Meng}]{liu_log2vec:_2019}
\bibinfo{author}{F.~Liu}, \bibinfo{author}{Y.~Wen}, \bibinfo{author}{D.~Zhang},
  \bibinfo{author}{X.~Jiang}, \bibinfo{author}{X.~Xing},
  \bibinfo{author}{D.~Meng},
\newblock \bibinfo{title}{{Log2Vec}: {A} {Heterogeneous} {Graph} {Embedding}
  {Based} {Approach} for {Detecting} {Cyber} {Threats} {Within} {Enterprise}},
\newblock in: \bibinfo{booktitle}{Proceedings of the 2019 {ACM} {SIGSAC}
  {Conference} on {Computer} and {Communications} {Security}}, {CCS} '19,
  \bibinfo{publisher}{ACM}, \bibinfo{address}{New York, NY, USA},
  \bibinfo{year}{2019}, pp. \bibinfo{pages}{1777--1794}. \URLprefix
  \url{http://doi.acm.org/10.1145/3319535.3363224}.
  \DOIprefix\doi{10.1145/3319535.3363224}, \bibinfo{note}{event-place: London,
  United Kingdom}.
\bibitem[{Wang et~al.(2020)Wang, Hassan, Li, Jee, Yu, Zou, Rhee, Chen, Cheng,
  Gunter, and Chen}]{wang_you_nodate}
\bibinfo{author}{Q.~Wang}, \bibinfo{author}{W.~U. Hassan},
  \bibinfo{author}{D.~Li}, \bibinfo{author}{K.~Jee}, \bibinfo{author}{X.~Yu},
  \bibinfo{author}{K.~Zou}, \bibinfo{author}{J.~Rhee},
  \bibinfo{author}{Z.~Chen}, \bibinfo{author}{W.~Cheng}, \bibinfo{author}{C.~A.
  Gunter}, \bibinfo{author}{H.~Chen},
\newblock \bibinfo{title}{You {Are} {What} {You} {Do}: {Hunting} {Stealthy}
  {Malware} via {Data} {Provenance} {Analysis}},
\newblock \bibinfo{year}{2020}, p.~\bibinfo{pages}{17}.
\bibitem[{Liu et~al.(2018)Liu, Zhang, Li, Jee, Li, Wu, Rhee, and
  Mittal}]{liu_towards_2018}
\bibinfo{author}{Y.~Liu}, \bibinfo{author}{M.~Zhang}, \bibinfo{author}{D.~Li},
  \bibinfo{author}{K.~Jee}, \bibinfo{author}{Z.~Li}, \bibinfo{author}{Z.~Wu},
  \bibinfo{author}{J.~Rhee}, \bibinfo{author}{P.~Mittal},
\newblock \bibinfo{title}{Towards a {Timely} {Causality} {Analysis} for
  {Enterprise} {Security}},
\newblock \bibinfo{publisher}{Internet Society}, \bibinfo{year}{2018}.
  \DOIprefix\doi{10.14722/ndss.2018.23254}.
\bibitem[{{Ul Hassan} et~al.(2020){Ul Hassan}, Li, Jee, Yu, Zou, Wang, Chen,
  Li, Gui, Bates, and Gui}]{UlHassan2020}
\bibinfo{author}{W.~{Ul Hassan}}, \bibinfo{author}{D.~Li},
  \bibinfo{author}{K.~Jee}, \bibinfo{author}{X.~Yu}, \bibinfo{author}{K.~Zou},
  \bibinfo{author}{D.~Wang}, \bibinfo{author}{Z.~Chen},
  \bibinfo{author}{Z.~Li}, \bibinfo{author}{J.~Gui},
  \bibinfo{author}{A.~Bates}, \bibinfo{author}{J.-i. Gui},
\newblock \bibinfo{title}{{ This is Why We Can't Cache Nice Things:
  Lightning-Fast Threat Hunting using Suspicion-Based Hierarchical Storage}},
\newblock in: \bibinfo{booktitle}{ACSAC 2020}, volume~\bibinfo{volume}{14},
  \bibinfo{publisher}{ACM}, \bibinfo{year}{2020}. \URLprefix
  \url{https://doi.org/10.1145/3427228.3427255}.
  \DOIprefix\doi{10.1145/3427228.3427255}.
\bibitem[{Jiang et~al.(2006)Jiang, Walters, Xu, Spafford, Buchholz, and
  Wang}]{jiang_provenance-aware_2006}
\bibinfo{author}{X.~Jiang}, \bibinfo{author}{A.~Walters},
  \bibinfo{author}{D.~Xu}, \bibinfo{author}{E.~H. Spafford},
  \bibinfo{author}{F.~Buchholz}, \bibinfo{author}{Y.-M. Wang},
\newblock \bibinfo{title}{Provenance-{Aware} {Tracing} of {Worm} {Break}-in and
  {Contaminations}: {A} {Process} {Coloring} {Approach}},
\newblock in: \bibinfo{booktitle}{26th {IEEE} {International} {Conference} on
  {Distributed} {Computing} {Systems} ({ICDCS}'06)}, \bibinfo{year}{2006}, pp.
  \bibinfo{pages}{38--38}. \DOIprefix\doi{10.1109/ICDCS.2006.69}.
\bibitem[{Hossain et~al.(2020)Hossain, Sheikhi, and Sekar}]{hossaincombating}
\bibinfo{author}{M.~N. Hossain}, \bibinfo{author}{S.~Sheikhi},
  \bibinfo{author}{R.~Sekar},
\newblock \bibinfo{title}{Combating dependence explosion in forensic analysis
  using alternative tag propagation semantics},
\newblock in: \bibinfo{booktitle}{IEEE S\&P 2020}, \bibinfo{year}{2020}.
\bibitem[{De~Nardo et~al.(2008)De~Nardo, Ranzato, and Tapparo}]{de2008subgraph}
\bibinfo{author}{L.~De~Nardo}, \bibinfo{author}{F.~Ranzato},
  \bibinfo{author}{F.~Tapparo},
\newblock \bibinfo{title}{The subgraph similarity problem},
\newblock \bibinfo{journal}{IEEE Transactions on Knowledge and Data
  Engineering} \bibinfo{volume}{21} (\bibinfo{year}{2008})
  \bibinfo{pages}{748--749}.
\bibitem[{Yan et~al.(2006)Yan, Xu, Zhang, Zhang, Yang, and Lin}]{yan2006graph}
\bibinfo{author}{S.~Yan}, \bibinfo{author}{D.~Xu}, \bibinfo{author}{B.~Zhang},
  \bibinfo{author}{H.-J. Zhang}, \bibinfo{author}{Q.~Yang},
  \bibinfo{author}{S.~Lin},
\newblock \bibinfo{title}{Graph embedding and extensions: A general framework
  for dimensionality reduction},
\newblock \bibinfo{journal}{IEEE transactions on pattern analysis and machine
  intelligence} \bibinfo{volume}{29} (\bibinfo{year}{2006})
  \bibinfo{pages}{40--51}.
\bibitem[{Wang et~al.(2014)Wang, Zhang, Feng, and Chen}]{wang2014knowledge}
\bibinfo{author}{Z.~Wang}, \bibinfo{author}{J.~Zhang},
  \bibinfo{author}{J.~Feng}, \bibinfo{author}{Z.~Chen},
\newblock \bibinfo{title}{Knowledge graph embedding by translating on
  hyperplanes},
\newblock in: \bibinfo{booktitle}{Twenty-Eighth AAAI conference on artificial
  intelligence}, \bibinfo{year}{2014}.
\bibitem[{Goyal and Ferrara(2018)}]{goyal2018graph}
\bibinfo{author}{P.~Goyal}, \bibinfo{author}{E.~Ferrara},
\newblock \bibinfo{title}{Graph embedding techniques, applications, and
  performance: A survey},
\newblock \bibinfo{journal}{Knowledge-Based Systems} \bibinfo{volume}{151}
  (\bibinfo{year}{2018}) \bibinfo{pages}{78--94}.
\bibitem[{git(2020)}]{gitpwnd}
\bibinfo{title}{nccgroup/gitpwnd: Gitpwnd is a network penetration tool that
  lets you use a git repo for command and control of compromised machines},
  \bibinfo{year}{2020}. \URLprefix \url{https://github.com/nccgroup/gitpwnd}.

\end{thebibliography}

\end{document}